\documentclass[epj]{svjour}

\usepackage{latexsym}
\usepackage{amsmath}
\usepackage{amssymb}
\usepackage{phonetic}

\usepackage{epsfig}
\usepackage{graphicx}
\usepackage{graphics}

\numberwithin{equation}{section}

\newcommand{\be}{\begin{equation}}
\newcommand{\ee}{\end{equation}}
\newcommand{\bea}{\begin{eqnarray}}
\newcommand{\eea}{\end{eqnarray}}
\newcommand{\bra}{\langle}
\newcommand{\ket}{\rangle}

\newcommand{\im}{{\mathrm{Im}}}
\newcommand{\re}{{\mathrm{Re}}}
\newcommand{\Tr}{{\mathrm{Tr}}}
\newcommand{\tr}{{\mathrm{tr}}}

\newcommand{\xv}{{\mathbf x}}

\newcommand{\vecx}{{\mathbf x}}

\newcommand{\half}{\frac{1}{2}}

\newcommand{\eps}{\epsilon}

\newcommand{\om}{\omega}
\newcommand{\bean}{\begin{eqnarray*}}
\newcommand{\eean}{\end{eqnarray*}}
\newcommand{\nn}{\nonumber}
\newcommand{\rmR}{{\rm R}}
\newcommand{\rmI}{{\rm I}}
\newcommand{\id}{1\!\!1}
\newcommand{\hm}{\hspace*{-0.15cm}}

\newcommand{\cP}{{\cal P}}
\newcommand{\e}{{\rm e}}
\newcommand{\grad}{{\rm grad \, }}

\newcommand{\dd}{\mbox{\hausad}}

\begin{document}

\title{Controlling complex Langevin dynamics at finite density}

\author{Gert Aarts\inst{1}\thanks{g.aarts@swan.ac.uk}
\and Lorenzo Bongiovanni\inst{1}\thanks{pylb@swan.ac.uk}
\and Erhard Seiler\inst{2}\thanks{ehs@mppmu.mpg.de}
\and D\'enes Sexty\inst{3}\thanks{d.sexty@thphys.uni-heidelberg.de}
\and Ion-Olimpiu Stamatescu\inst{3,4}\thanks{i.o.stamatescu@thphys.uni-heidelberg.de}
}

\institute{ 
Department of Physics, College of Science, Swansea University, Swansea, United Kingdom
\and 
Max-Planck-Institut f\"ur Physik (Werner-Heisenberg-Institut), M\"unchen, Germany
\and
Institut f\"ur Theoretische Physik, Universit\"at Heidelberg, Heidelberg, Germany
\and
FEST, Heidelberg, Germany
}

\date{Received: date / Revised version: date}

\abstract{
At nonzero chemical potential the numerical sign problem in lattice field theory limits the use of standard algorithms based on importance sampling. 
Complex Langevin dynamics provides a possible solution, but it has to be applied with care.
In this review, we first summarise our current understanding of the approach, combining analytical and numerical insight.
In the second part we study SL($N, \mathbb{C}$) gauge cooling, which was introduced recently as a tool to control complex Langevin dynamics in nonabelian gauge theories. We present new results in Polyakov chain models and in QCD with heavy quarks and compare various adaptive cooling implementations.
}

\maketitle


\section{Introduction}
\label{sec:intro}

One of the outstanding questions in the theory of the strong interactions, Quantum Chromodynamics, is the determination of its phase diagram in the plane of temperature and baryon chemical potential. Nonperturbative studies using lattice QCD are hindered by the presence of the complex fermion determinant and the resulting sign problem: since the Boltzmann weight in the partition function is complex, standard numerical approaches relying on importance sampling are only of limited use \cite{deForcrand:2010ys}. As a result various alternative approaches in numerical lattice field theory have been developed during the past years; a comprehensive overview can be found in Ref.\ \cite{Aarts:2013bla}.

In this review we focus on the complex Langevin (CL) method, in which a {\em complexified} configuration space is explored stochastically and the presence of a positive weight is not required \cite{Parisi:1984cs,Klauder:1983}. In fact, due to the complex nature of the Boltzmann weight, the dynamics naturally takes place in this larger complexified configuration space, and it is precisely this extension that allows for a solution of the sign problem. 
While it has been successfully demonstrated that the sign problem can be solved in a number of theories, including those in which the sign problem is severe \cite{Aarts:2008rr,Aarts:2008wh,arXiv:1006.0332,Aarts:2011zn}, success is not guaranteed \cite{Ambjorn:1985iw,Ambjorn:1986fz,Berges:2006xc,Berges:2007nr,arXiv:0912.3360,arXiv:1005.3468,arXiv:1101.3270,Pawlowski:2013pje}. 
The first aim of this contribution is to summarise why problems may arise and how they can be detected in practice. 
Recently, definite progress has been made in nonabelian gauge theories, by employing gauge cooling during the complex Langevin evolution \cite{Seiler:2012wz}.
The second aim of this paper is to discuss this in some detail and present new results comparing various gauge cooling  implementations.

This contribution is organised as follows: in the following section, we contrast complex and real Langevin dynamics, and remind the reader why the standard proof for the justification of real Langevin dynamics is no longer applicable. A justification following a different route is given in Sec.\ \ref{sec:dist}: we identify criteria for correctness and  emphasise the importance of understanding the distribution in the complexified space. 
This is illustrated in Sec.\ \ref{sec:su3}, where two three-dimensional models are contrasted: the SU(3) spin model and the XY model, both at nonzero chemical potential. While for the former CL works in the entire phase diagram, for the latter this is only true in the ordered part of the phase diagram.
In Sec.\ \ref{sec:cool} we come to nonabelian dynamics and explain how gauge cooling can be used to control the distribution in the complexified space, using Polyakov chain models and QCD with heavy quarks  for  illustration. It is shown that a judicious choice of cooling implementations leads to well-controlled dynamics in SL($N, \mathbb{C}$).
Ref.~\cite{Aarts:2013bla} reviews CL with an emphasis on aspects complementary to the ones discussed here.

\section{Real versus complex Langevin dynamics}
\label{sec:real}

In order to indicate the difference in theoretical status between real and complex Langevin dynamics, we start with a brief reminder, emphasising why the ``standard proofs'' of correctness are no longer available in the case of a complex action. The results in this section are well known \cite{Damgaard:1987rr}. 

For simplicity, we consider one degree of freedom $x$, but the expressions can easily be generalized. 
We start with the partition function
\be
\label{eq:z}
Z =  \int dx\, e^{-S(x)}, \quad\quad S(x)\in \mathbb{R}. 
\ee
The corresponding Langevin equation reads
\be
\dot x  = -\partial_xS(x)+\eta,
\ee
where the random noise $\eta$ is Gaussian distributed, with $\bra\eta(t)\eta(t')\ket=  2\delta(t-t')$.
The evolution of expectation values of observables $O(x)$ in Langevin time is represented, after averaging over the noise, by the evolution of the associated distribution $\rho(x,t)$, according to
\be
\bra O\ket_{\rho(t)} = \int dx\, \rho(x,t)O(x).
\ee
Given the Langevin equation for $x(t)$, it follows that $\rho(x,t)$ satisfies a Fokker-Planck (FP) equation,
\be
\label{eq:rho}
\dot \rho(x,t) = \partial_x\left(\partial_x+S'(x)\right)\rho(x,t).
\ee
This FP equation has a stationary solution $\rho(x) \sim e^{-S(x)}$, which is the first prerequisite. Moreover, it turns out that this solution is typically reached exponentially fast. In order to see this, write $\rho(x,t) = \psi(x,t) e^{-\half S(x)}$; it then follows that $\psi(x,t)$ satisfies a Schr\"odinger-like equation, 
\be
\dot\psi(x,t) = -H_{\rm FP}\psi(x,t),
\ee
with the FP Hamiltonian,  
\be
\label{eq:HFP}
H_{\rm FP} = Q^\sharp Q,
\ee 
in which
\be
Q^\sharp=  -\partial_x +\half S'(x),
\quad\quad\quad
Q =  \partial_x +\half S'(x).
\ee
Since $Q^\sharp = Q^\dagger$, the FP Hamiltonian is semi-positive definite:
if $|S'|$ goes to infinity as $|x|$ does, it has a discrete spectrum and a nondegenerate 
ground state (cf. Refs.\ \cite{reed-simon,Damgaard:1987rr}). Let us denote the eigenvalues
of $H_{\rm FP}$ by $\lambda\geq 0$, with $\lambda=0$ corresponding to 
\be
Q\psi(x)=0 \quad\quad\Leftrightarrow\quad\quad \psi(x) \sim e^{-\half S(x)},
\ee
then 
\be
\psi(x,t) = c_0 e^{-\half S(x)} + \sum_{\lambda>0} c_\lambda e^{-\lambda t}  \to c_0 e^{-\half S(x)},
\ee
and the correct distribution is reached exponentially fast.

In the case of a complex action, $S(x)\in \mathbb{C}$, and assuming that $S$ can be analytically continued into the complex plane, one can attempt to follow the same line of reasoning. 
 Replacing $x$ by $z$ to emphasise that the dynamics is now complex, the Langevin equation 
\be
\dot z  = -\partial_zS(z)+\eta
\ee
and FP equation
\be
\label{eq:FPrho}
\dot \rho(z,t) = \partial_z\left(\partial_z+S'(z)\right)\rho(z,t)
\ee
can still be written down. However, since in this case $\rho(z,t)$ is complex-valued, it is not a probability distribution. Moreover, the associated FP Hamiltonian (\ref{eq:HFP}) is no longer semi-positive definite, since $Q^\sharp \neq Q^\dagger$.

In order to discuss the (real and positive) probability distribution in the complex plane, we write explicitly $x\to z=x+iy$ and consider the 
complex Langevin equations,
\be
\dot x =  K_x +  \eta_\rmR, 
\quad\quad\quad
\dot y =  K_y +  \eta_\rmI, 
\ee
with the drift terms 
\be
K_x =  -\re\,\partial_z S(z),
\quad\quad
K_y = -\im\,\partial_z S(z).
\ee
Here we momentarily introduced ``complex'' noise, writing $\eta=\eta_\rmR+i\eta_\rmI$ \cite{arXiv:0912.3360,arXiv:1101.3270}. Starting from $\bra\eta(t)\eta(t')\ket=  2\delta(t-t')$, it follows that the real and imaginary components of the noise satisfy
\bea
 \bra\eta_\rmR(t)\eta_\rmR(t') \ket =&&\hm  2N_\rmR\delta(t-t'), 
\nn \\
 \bra \eta_\rmI(t)\eta_\rmI(t')\ket =&&\hm  2N_\rmI\delta(t-t'), 
 \\
\nn
\bra\eta_\rmR(t)\eta_\rmI(t') \ket =&&\hm 0,
\eea
with $N_\rmR-N_\rmI=1$ and $N_\rmI\geq 0$. 
 The evolution of the expectation value of holomorphic observables $O(x)$ is now represented by
\be
\bra O \ket_{P(t)} = \int dxdy\, P(x,y;t) O(x+iy),
\ee 
where the associated distribution $P(x,y;t)$ satisfies the (real) FP equation
\bea
\nn
\dot P(x,y;t) = &&\hm \big[ \partial_x\left(N_\rmR\partial_x - K_x\right) \\
&&\hm
+\partial_y \left(N_\rmI\partial_y - K_y\right) \big] P(x,y;t).
\label{eq:FPreal}
\eea
Although this equation is superficially of the same form as Eq.\ (\ref{eq:rho}), with simply twice as many degrees of freedom, it is in fact much harder to solve. In particular there are no generic solutions known, even in the stationary case. Some recent solutions in special cases can be found in Refs.~\cite{Aarts:2009hn,arXiv:0912.3360,Duncan:2012tc}. 
Moreover, most likely it is not possible to express the dynamics in terms of a semi-positive definite FP Hamiltonian, so that convergence to a stationary solution is not guaranteed. The best one may hope is a Hamiltonian whose spectrum is in the right half plane, but this should be sufficient.
 Nevertheless, in many cases of interest  a stationary distribution emerges when the CL equations are solved numerically: one goal of this review is to indicate how properties of this distribution relate to the reliability and justification of the approach.

We end this section with a description of how the CL equations appear in lattice field theory.
Consider first a real scalar field $\phi_{ax}$, where $a$ denotes additional internal degrees of freedom. 
After discretising Langevin time with time step $\eps$ and using the lowest-order discretisation scheme,  the Langevin equation takes the form
\be
\phi_{ax}(n+1) = \phi_{ax}(n) +\eps K_{ax}(n) +\sqrt\eps\eta_{ax}(n),
\ee
with the drift 
\be
K_{ax} =  -\delta S[\phi]/\delta \phi_{ax},
\ee
and the noise (we only consider real noise from now on, $N_\rmI=0$) satisfying
$\bra \eta_{ax}(n)\eta_{a'x'}(n')\ket = 2\delta_{aa'}\delta_{xx'}\delta_{nn'}$.
When the drift is complex, the field is complexified, $\phi_{ax} \to \phi_{ax}^\rmR+i\phi_{ax}^\rmI$, as above.

For gauge or matrix theories, with gauge group SU($N$),  the dynamics is written for group elements, living on links,  as 
\cite{Batrouni:1985jn,Berges:2006xc}
\be
 U_{x,\nu}(n+1)  = R_{x,\nu}(n)\, U_{x,\nu}(n),
 \ee
with
\be
\label{eq:R}
R_{x,\nu} = \exp \left[ i\sum_a\lambda_a\left( \eps K_{ax\nu} +\sqrt \eps \eta_{ax\nu}\right) \right].
\ee
Here $\lambda_a$ ($a=1,\ldots N^2-1$)  are the traceless, hermitian Gell-Mann matrices, normalized as $\Tr\,\lambda_a\lambda_b=2\delta_{ab}$, and the drift is determined by 
\be
K_{ax\nu} =  -D_{ax\nu} S[U],
\ee
where the action $S[U]$ may contain a contribution from the fermion determinant as $S_F = - \ln\det M$.
The differential operator $D$ is defined as
\be
D_{ax\nu} f(U) = \partial_\alpha f\left( e^{i\alpha\lambda_a}U_{x,\nu}\right) \Big|_{\alpha=0}.
\ee
In the case that the action is complex, it is easy to see that the drift
$K\neq K^\dagger$, and hence $R^\dagger R\neq \id$, although its determinant $\det R=1$. In SU($N$) theories, one finds therefore that $U\in$ SL($N$,$\mathbb{C}$) during the CL process.

Finally, we note that in practice it may be necessary to choose the stepsize $\eps$ adaptively, to avoid instabilities due to runaway solutions \cite{arXiv:0912.0617}. In general, discretisations of the Langevin equation result in finite stepsize corrections, which are linear in $\eps$ for the Euler scheme given above. This can be improved by employing higher-order discretisation schemes \cite{CCC,Petersen:1996by,Aarts:2011zn}.

\begin{figure}[t]
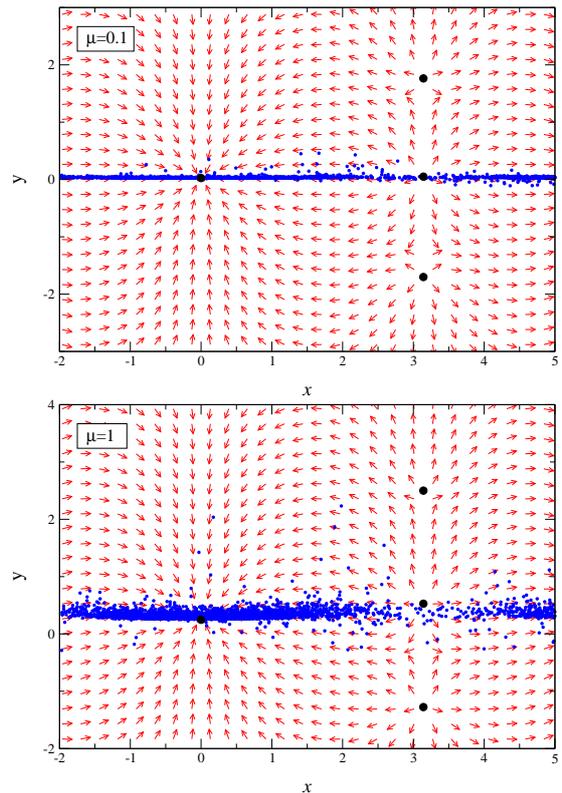

  \begin{center}
   \vspace*{0.1cm}
    \includegraphics[width=0.4\textwidth]{plot_full_b1_k0.25_m0.1_scat.eps}
    \includegraphics[width=0.4\textwidth]{plot_full_b1_k0.25_m1_scat.eps}
     \end{center}
  \caption{U(1) one-link model: scatter plots in the complex plane, for $\mu=0.1$ and 1, at fixed $\beta=1$ and $\kappa=1/2$ \cite{Aarts:2008rr}.}
  \label{fig:scat1}
\end{figure}

\begin{figure}[t]
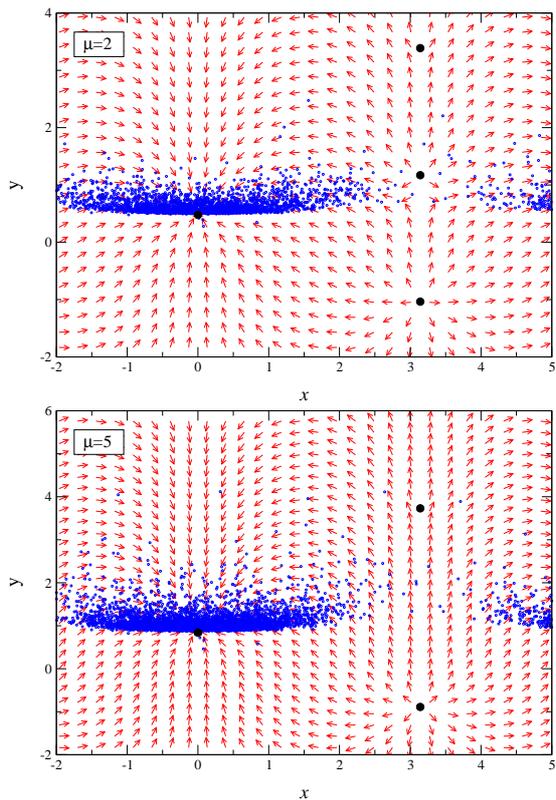

  \begin{center}
      \includegraphics[width=0.4\textwidth]{plot_full_b1_k0.25_m2_scat.eps}
    \includegraphics[width=0.4\textwidth]{plot_full_b1_k0.25_m5_scat.eps}
  \end{center}
  \caption{As in Fig.\ \ref{fig:scat1}, for $\mu=2$ and 5.}
  \label{fig:scat2}
\end{figure}

\section{Distributions in the complex plane}
\label{sec:dist}

A crucial role is played by the distribution $P(x,y)$ in the enlarged configuration space (we drop the $t$ dependence to indicate that  from now on we only consider the stationary distribution).
While it is not guaranteed that a stationary solution exists in general, in many interesting cases we find that it does and that it can be constructed by brute force, namely by solving the CL equations and collecting the coordinates visited during the stochastic process. Properties of the distribution can then be   studied using histograms and scatter plots.
 This method is well suited for zero-dimensional models; when more degrees of freedom are involved, it is useful to integrate out most coordinates and focus on the ones of interest. 

To illustrate this, we consider a simple U(1) one-link model  \cite{Aarts:2008rr}, with the partition function
\be
\label{eq:u1}
Z = \int_{-\pi}^\pi dx\, e^{\beta\cos x} \det M(x;\mu),
\ee
with
\be
\det M(x;\mu) = 1+\kappa\cos(x-i\mu).
\ee
Its structure is inspired by QCD in the hopping expansion, with a complex ``determinant'', satisfying $[\det M(x;\mu)]^*=\det M(x,-\mu^*)$. When $\mu=0$, the determinant is  real and the Langevin process takes place on the real axis, $y=0$ (we take $|\kappa|<1$, so that the determinant is positive when $\mu=0$).
 Increasing $\mu$ results in the trajectories presented in Figs.\ \ref{fig:scat1}, \ref{fig:scat2}. 
Shown here are CL trajectories, indicated by the small (blue) dots in the $x-y$ plane (note that the dots are separated by 500 updates). The arrows indicate the direction of the classical forces, while the larger (black) dots are classical fixed points, where the drift terms vanish, $K_x=K_y=0$.
A small increase in $\mu$ moves the distribution slightly into the complex plane, see Fig.\ \ref{fig:scat1} (top). Increasing $\mu$ even further, as in Fig.\ \ref{fig:scat2}, moves the distribution further out, but it remains localised and bound to the attractive fixed point at $x=0$. Note that the distribution is periodic in the $x$ direction and highly asymmetric in the $y$ direction. Expectation values of observables obtained with CL agree with the exact results, which can be computed in this case \cite{Aarts:2008rr}.

However, in most cases of interest exact results are not available and CL has to be 
justified without recourse to a comparison. For this a good understanding of the 
properties of the distribution is necessary. We found that it is essential for the 
mathematical justification of the approach that the distribution is well localised in the 
complexified configuration space and in particular shows a fast decay in the imaginary 
directions. The argument goes along the following lines 
\cite{arXiv:0912.3360,arXiv:1101.3270}. We consider the two expectation values, 
\bea 
&& \bra O\ket_{\rho(t)} = \int dz\, \rho(z,t)O(z), \\ 
&&\bra O \ket_{P(t)} = \int dxdy\, P(x,y;t) O(x+iy), 
\eea 
in which $\rho(z,t)$ and $P(x,y;t)$ were introduced above and
satisfy their respective FP equations, (\ref{eq:FPrho}) and (\ref{eq:FPreal}). The goal is to show that 
\be 
\bra O\ket_{\rho(t)} = \bra O\ket_{P(t)}, 
\ee 
for all times or possibly only in the limit that $t\to\infty$, 
starting from identical initial conditions (although the initial condition dependence 
is expected to drop out).  This can indeed be shown by moving the time evolution from 
the densities to the observables, making use of the Cauchy-Riemann equations and 
performing integration by parts, assuming that no boundary terms at infinity are 
produced. Crucial is therefore the decay at $y\to\pm\infty$ of the product 
$P(x,y)O(x+iy)$ for a suitable class of observables. In Ref.~\cite{arXiv:1101.3270} we 
showed concretely for a toy model how the failure of the CL method is associated with 
insufficient decay of $P(x,y)O(x+iy)$.

In Ref.\ \cite{arXiv:1101.3270} we also identified certain {\it consistency conditions} 
which are necessary for the CL method to give correct results. They involve the 
complex ``Langevin operator''  
\be
 \tilde L \equiv \left[\partial_z-(\partial_z S(z))\right]  \partial_z,
\ee
which governs the evolution of holomorphic observables; the consistency conditions are 
\be
\bra \tilde L O \ket=0
\ee
for all observables $O$. Here the expectation value is taken with respect to the equilibrium 
distribution $P(x,y)$, or equivalently after averaging over the noise.
 Formally these conditions just express the stationarity of the 
equilibrium expectation values, but actually their derivation involves integrations by 
part and therefore requires sufficient decay of the equilibrium distribution. If the 
decay is insufficient, the conditions may fail either by being nonzero, by diverging, 
or by being ill-defined (reflected in huge fluctuations). While these conditions are 
only necessary, in principle they become sufficient if they hold for a sufficiently 
large set (a basis in some sense) of observables \cite{arXiv:1101.3270}. They were 
also shown to be formally equivalent to the Schwinger-Dyson equations.
In Ref.\ \cite{arXiv:1101.3270} the strength of these conditions, applied to a 
few simple observables, was tested and generally we found that they are successful in 
weeding out incorrect results.

Another intuitive insight relates to the stability of the real manifold (i.e.\ $y=0$) under small complex fluctuations \cite{Aarts:2012ft}. In the example discussed above, the stationary distribution at $\mu=0$ is given by $P(x,y;\mu=0) = \rho(x;\mu=0)\delta(y)$. In Fig.~\ref{fig:scat1} (top) it can be seen that turning on $\mu$, and thereby introducing a nonzero force in the imaginary direction, does not drastically change the distribution: it only slightly widens in the $y$ direction. The distribution at  $\mu\sim 0$ is therefore connected to the one at $\mu=0$.
Although this condition is neither sufficient nor necessary (there could be another, much wider distribution which gives the correct result), it helps in interpreting results in practice, as we will demonstrate below.

Finally, we note that when complex noise is used ($N_\rmI>0$), there is additional diffusion in the imaginary direction. This has the effect of widening the distribution. We found that in most cases this has a detrimental effect on the results of CL  \cite{arXiv:0912.3360,arXiv:1101.3270}. In general complex noise is therefore not recommended, although in some cases a small amount of complex noise ($N_\rmI\ll 1$) can be used to regulate numerical solutions of the FP equation (\ref{eq:FPreal}) \cite{arXiv:0912.3360}.

\section{SU(3) spin model vs XY model}
\label{sec:su3}

In this section we apply these ideas to contrast two three-dimensional models, the SU(3) spin model and the XY model, both at nonzero chemical potential. It turns out that CL works very well in the SU(3) spin model \cite{Aarts:2011zn}, but that it fails in part of the phase diagram of the XY model \cite{arXiv:1005.3468}. Understanding the difference yields considerable insight into the approach \cite{Aarts:2012ft}.
We note that both can also be solved with worldline/flux methods, which is sign-problem free, so that ``exact'' results are available \cite{arXiv:1001.3648,arXiv:1104.2503,Mercado:2012ue}.

The three-dimensional SU(3) spin model is an effective Polyakov loop model for  QCD with heavy quarks, obtained using a combined strong-coupling/hopping expansion.
The action reads
\bea
\nn
 S  =&&\hm  -\beta\sum_{x,\nu} \left[  \Tr\,U_x \Tr\,U^{-1}_{x+\hat\nu} +  \Tr\,U_x^{-1} \Tr\,U_{x+\hat\nu}\right]  \\
 &&\hm  -h \sum_x\left[ e^\mu \Tr\,U_x + e^{-\mu} \Tr\,U_x^{-1} \right],
\eea
where the $U_x$'s are SU(3) matrices, living on a three-dimensional lattice with $\Omega=N^3$ sites, and the additional sum in the first line is over the nearest neighbours.
Static quarks are represented by Polyakov loops and couple to the chemical potential in the usual way, resulting in a complex action, 
$S^*(\mu) = S(-\mu^*)$. This model has been solved with CL in the classic papers \cite{KW,BGS}; here we show results from Ref.\ \cite{Aarts:2011zn}.
The flux representation is discussed in Refs.\ \cite{arXiv:1104.2503,Mercado:2012ue}, while a recent mean-field study appears in Ref.\ \cite{Greensite:2012xv}. The model is part of a family of 
strong-coupling/heavy-quark effective models, constructed in Refs.\ \cite{Fromm:2011qi,Fromm:2012eb}. We note that in the latter CL has been used  to study the phase structure.

\begin{figure}[t]
  \centerline{
    \includegraphics[width=0.45\textwidth]{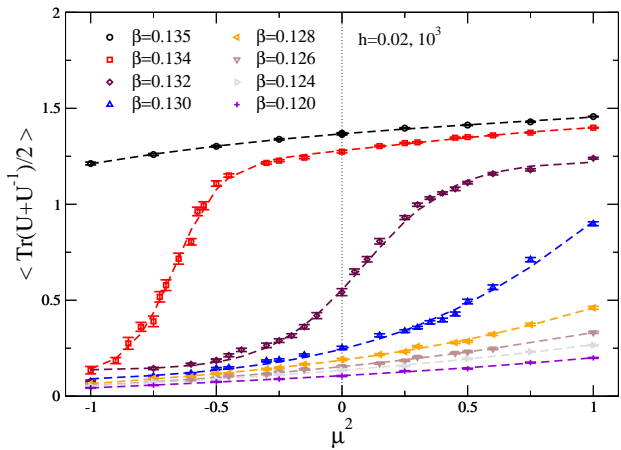}
  }
  \caption{SU(3) spin model: $\bra \Tr (U+U^{-1})/2\ket$ as a function of $\mu^2$, for several values of $\beta$ on a $10^3$ lattice \cite{Aarts:2011zn}.}
  \label{fig:su3}
\end{figure}

The action for the three-dimensional XY model is
\bea
\nn
S = &&\hm -\half\beta\sum_{x,\nu} \left[ e^{\mu \delta_{\nu,0}}  U_xU^{-1}_{x+\hat \nu} + e^{-\mu \delta_{\nu,0} } U_x^{-1} U_{x+\hat\nu} \right]
\\
= &&\hm -\beta\sum_{x,\nu}  \cos\left(\phi_x-\phi_{x+\hat \nu}-i\mu\delta_{\nu,0} \right).
\eea
Here the degrees of freedom are U(1) phases, $U_x=e^{i\phi_x}$. The chemical potential couples to 
the conserved Noe\-ther charge and the action is complex, as above. This model has been studied using a worldline formulation in Ref.~\cite{arXiv:1001.3648} and more recently in Ref.\ \cite{Langfeld:2013kno}.
Here we discuss the results from Ref.\  \cite{arXiv:1005.3468}.
Both models have a similar phase structure: a disordered phase at small $\beta$ and $\mu$ and an 
ordered phase at large $\beta$ and/or $\mu$. In the SU(3) spin model, the transition weakens as $\mu$ is increased and turns into a crossover. The critical couplings at $\mu=0$ are $\beta_c=0.133-0.137$ (depending on $h$) in the SU(3) spin model \cite{Aarts:2011zn,Mercado:2012ue}
and $\beta_c = 0.45421$ in the XY model \cite{Campostrini:2000iw}.

\begin{figure}[t]
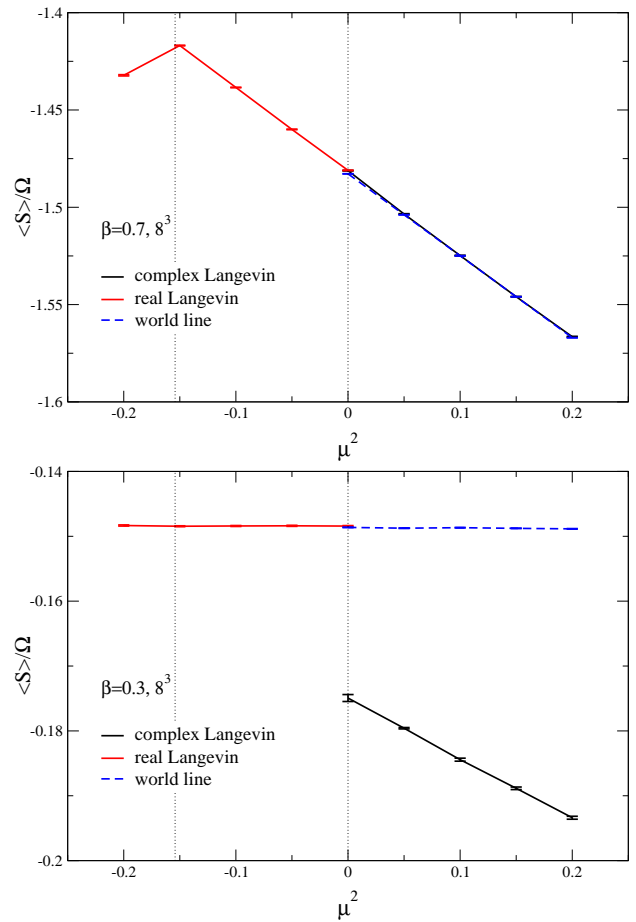

  \begin{center}
    \includegraphics[width=0.45\textwidth]{act_b0.7_x8.eps} \\
    \includegraphics[width=0.45\textwidth]{act_b0.3_x8.eps}
\end{center}  
  \caption{XY model: action density $\bra S\ket/\Omega$ as a function of $\mu^2$ in the ordered phase at  $\beta=0.7$ (top) and the disordered phase at $\beta=0.3$ (bottom) on an $8^3$ lattice. The dotted line at $\mu=i\pi/8$ indicates the Roberge-Weiss transition at imaginary $\mu$   \cite{arXiv:1005.3468}.}
  \label{fig:xy1}
\end{figure}

Out of the number of criteria we have developed to justify CL, we consider here (the lack of) analyticity in $\mu^2$ around $\mu^2=0$ and changes in the distribution in the complexified configuration space as $\mu$ is increased.
Note that negative $\mu^2$ corresponds to imaginary chemical potential, for which real Langevin dynamics can be used.  Fig.~\ref{fig:su3} shows  $\bra \Tr (U+U^{-1})/2\ket$ as a function of $\mu^2$ in the SU(3) spin model, for several values of $\beta$. It is clear that the transition from $\mu^2\leq 0$ (obtained with real Langevin) to $\mu^2>0$ (obtained with complex Langevin) is smooth and that the observable is analytic in $\mu^2$, as it should be.
This interpretation has been confirmed with simulations using the flux representation \cite{Mercado:2012ue}.

This should be contrasted with what happens in the XY model. In Fig.\ \ref{fig:xy1} we show the expectation value of the action density around $\mu^2\sim 0$, for $\beta=0.7$ in the ordered phase (top) and $\beta=0.3$ in the disordered phase (bottom).
While in the ordered phase the observable is smooth in $\mu^2$, in the disordered phase we observe a clear nonanalyticity, which is interpreted as a failure of CL. This conclusion is further supported by a comparison with results obtained in the world-line formulation at positive $\mu^2$: again we see agreement in the ordered phase but disagreement in the disordered phase. We also note that the data point at $\mu^2=0$ in the lower line of the bottom graph has been obtained using complex initial conditions for the Langevin process. The resulting disagreement with the expected result indicates that in this case the real manifold is unstable and that the  distribution at $\mu\sim 0$ is not connected to the one at $\mu=0$ (see below).

\begin{figure}[t]
 \vspace*{-0.8cm}
  \centerline{
    \includegraphics[width=0.5\textwidth]{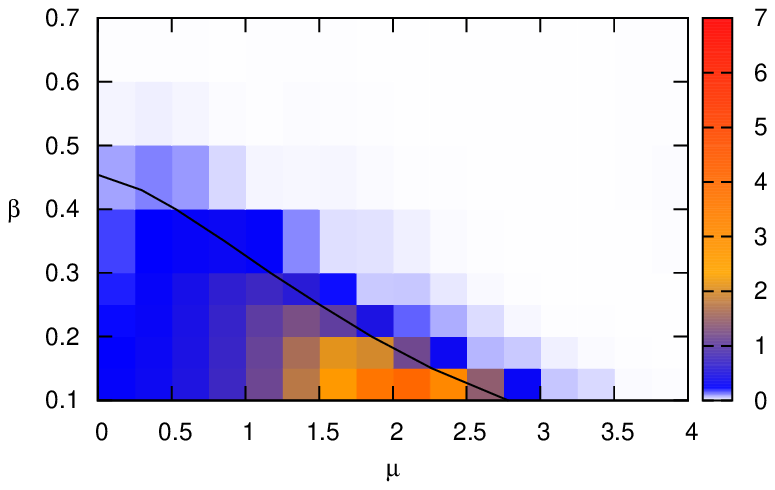}
  }
    \caption{XY model: 
    normalised difference between the CL and the worldline results for the action density,
    $\Delta S$, in the $\beta - \mu$ plane. The phase boundary between the ordered 
   and the disordered phase is indicated with the black line  \cite{arXiv:1005.3468}.}
  \label{fig:xy2}
\vspace*{0.2cm}
  \centerline{
    \includegraphics[width=0.45\textwidth]{img2-i2_x8.eps}
  }
  \caption{XY model: width of the distribution in the imaginary direction, $\bra(\Delta\phi_\rmI)^2\ket$, as a function of $\mu$ for several $\beta$ values \cite{arXiv:1005.3468}.}
  \label{fig:xy3}
\end{figure}

The failure in the XY model turns out to be closely related to its phase structure, as demonstrated in Fig.\ \ref{fig:xy2}. What is shown here is the normalised difference between the CL result and the worldline result for the action density,
\be
\Delta S = \frac{\bra S\ket_{\rm cl} - \bra S\ket_{\rm wl} }{\bra S\ket_{\rm wl}}.
\ee
The phase boundary, indicated with the black line, is taken from Ref.\ \cite{arXiv:1001.3648}. We conclude  that  incorrect 
results are obtained in the disordered/transition region, while CL is seen to work in the ordered phase.
As an aside, we remark that these findings are independent of the volume used and therefore independent of the severity of the sign problem.

The question is how the mismatch seen in Figs.\ \ref{fig:xy1} (bottom) and \ref{fig:xy2} manifests itself in the distribution in the complexified space, $P[\phi_\rmR, \phi_\rmI]$ . Since this is a high-dimensional example, it is not possible to plot the distribution directly, as in Figs.\  \ref{fig:scat1}, \ref{fig:scat2}. However,  we can determine how the width of the distribution in the imaginary direction, 
\be
\bra(\Delta\phi_\rmI)^2\ket = \bra \phi_\rmI^2\ket - \bra \phi_\rmI\ket^2,
\ee
depends on $\mu$ and $\beta$. This is shown in Fig.\ \ref{fig:xy3}. At $\mu=0$,
$\bra(\Delta\phi_\rmI)^2\ket=0$ provided that real initial conditions for the Langevin process are used.
At large $\beta$ the width increases smoothly from 0 as $\mu$ is increased. At smaller $\beta$ on the other hand, the width jumps discontinuously. Note that we used complex initial conditions, introducing complexity even at $\mu=0$. Alternatively, one can obtain this discontinuity using an infinitesimal but nonzero $\mu$. We also note that at large $\mu$, deep in the ordered phase, the width decreases and the dynamics becomes effectively real, signalling that the sign problem is effectively absent \cite{Aarts:2012ft}.
We conclude therefore that in the disordered phase the distribution $P[\phi_\rmR,\phi_\rmI]$ at $\mu\sim0$ is not connected to the distribution $\rho[\phi]$ at $\mu=0$. The distribution $P[\phi_\rmR,\phi_\rmI]$ is broad and not sufficiently localised, leading to a breakdown of the validity of CL dynamics.
 
The difference between the SU(3) spin model and the XY model can be understood from an analysis of classical flow diagrams, as in Figs.\  \ref{fig:scat1}, \ref{fig:scat2}, after reducing them to effective one-link models \cite{Aarts:2012ft}. The crucial difference between the (abelian) XY model and the (nonabelian) SU(3) model is the presence of the reduced Haar measure in the latter. It is exactly this contribution that stabilises the dynamics, since it is purely restoring and directed towards attractive fixed points on the real manifold. On the other hand,  in the XY model there is no restoring component and the real manifold is unstable against small complex fluctuations \cite{Aarts:2012ft}.

\section{Gauge cooling}
\label{sec:cool}

The lessons from the previous sections are that the distribution in the complexified space should be sufficiently localised and that it is desirable that turning on the chemical potential leads to a smooth deformation of the initial distribution into the complexified configuration space. We now apply these insights to SU($N$) gauge theories.
As described at the end of Sec.\ \ref{sec:real}, during a CL process links $U$ originally in SU($N$) will naturally take values in the noncompact group SL($N, \mathbb{C}$), in which $U^\dagger\neq U^{-1}$. Before making the analytic continuation, we therefore first express the action and observables in terms of the holomorphic variables $U$ and $U^{-1}$, rather than in terms of $U^\dagger$.
 To determine to which extent SL($N, \mathbb{C}$) is explored and the distribution is localised in this larger group, a measure needs to be defined that signifies the distance from the real submanifold, i.e.\ from SU($N$). There are a number of possibilities, here we consider the ``distance'' \cite{Aarts:2008rr}
\be
\label{eq:d1}
\dd \equiv \frac{1}{N}\Tr\left(UU^\dagger - \id\right) \geq 0,
\ee
and the ``unitarity norm''
\be
\label{eq:norm}
\mbox{norm} \equiv \Tr \left(UU^\dagger - \id \right)^2 \geq 0.
\ee
In both cases equality is reached for unitary matrices. The inequality in Eq.\ (\ref{eq:d1})
 follows from the polar decomposition of $U\in$ SL($N, \mathbb{C}$), which states that $U = PV$, with $V\in$ SU($N$) and $P=P^\dagger$ a positive-definite  matrix with determinant $\det P=1$.

We demonstrate the approach using Polyakov chain models: $N_\ell$ links $U_k$ form a one-dimensional chain, with periodic boundary conditions, and the partition function reads
\be
Z = \int \prod_{k=1}^{N_\ell} dU_k\, e^{-S[U]}.
\ee 
The action, with complex coupling $\beta$, reads
\be
\label{eq:su2}
S = -\frac{\beta}{2} \Tr\left(U_1U_2\ldots U_{N_\ell} \right)
\ee
for SU(2)
and 
\be
\label{eq:actsu3}
S=  -\beta_1\Tr(U_1\ldots U_{N_\ell}) - \beta_2 \Tr(U_{N_\ell}^{-1} \ldots U_1^{-1})
\ee
for SU(3),  with
\bea
&& \beta_1(\mu) = \beta+\kappa e^\mu, 
\nn \\
&&\beta_2(\mu) =   \beta^*+\kappa e^{-\mu} = [\beta_1(-\mu^*)]^*.
\eea
The structure of these models is motivated by the effective one-link models encountered in QCD in the heavy-quark/strong-coupling limit  \cite{Aarts:2012ft}. In the case of SU(2), there is no sign problem at nonzero chemical potential, but a  complex-action problem appears at nonzero theta angle or in real time. The coupling in Eq.\ (\ref{eq:su2}) is chosen to be  complex to mimick this.

As in four-dimensional SU($N$) gauge theory there is a notion of gauge symmetry: a link starting at site $k$ transforms as
\be
\label{eq:gt}
U_k\to \Omega_k U_k \Omega_{k+1}^{-1}, 
\quad\quad\quad
\Omega_k = e^{i\lambda_a\om_a^k},
\ee
where here and below the sum is over $a=1,\ldots, N^2-1$.
Of course, these models are gauge-equivalent to one-link models (i.e.\ with $N_\ell=1$) and exact results can be easily computed by integrating over the reduced Haar measure \cite{Aarts:2012ft}. However, increasing $N_\ell$ offers the possibility to study how the performance of the CL algorithm depends on many gauge degrees of freedom, mimicking the four-dimensional case. 

During the CL simulation, a distribution $P[U, U^\dagger]$ is effectively being sampled (we explicitly indicated here that $P$ depends both on $U$ and $U^\dagger$). In the case that the outcome of the CL simulation is incorrect, we may now exploit the gauge freedom to attempt to change the distribution $P[U, U^\dagger]$ in SL($N, \mathbb{C}$). Note that an SL($N, \mathbb{C}$) gauge transformation takes the same form as in Eq.\ (\ref{eq:gt}), but with $\om_a \in \mathbb{C}$.
The distance $\dd$ and the unitarity norm are not invariant under this transformation, which offers the possibility to bring them closer to 0 by a suitable choice of the gauge parameters $\om_a$. Hence changes in $P[U,  U^\dagger]$ can be quantified by changes in  $\dd$ and the unitarity norm.

One may wonder why by gauge transformations one can improve the result for 
gauge invariant observables. We now answer several aspects of this question.
First one should remember that the 
requirement to obtain correct expectation values for {\it holomorphic} 
observables does not determine the distribution in the complexified space; 
this has been discussed in some detail in the appendix of Ref.\ \cite{Aarts:2012ft}. 
In principle there is therefore tremendous freedom to select 
a process having good falloff properties at infinity, which can then
be sampled efficiently.
Gauge invariant holomorphic observables allow in addition changing the 
process by using complexified gauge transformations. This also changes the 
process and can be used to prevent it from moving out too far along the 
gauge orbits, which are noncompact due to complexification.  As will be 
seen, this is in fact necessary in order to keep numerical errors from 
building up and leading to incorrect or unstable results.

We also note that the procedure of ``gauge cooling", to be introduced shortly, 
 superficially seems to make the process `less ergodic', 
because it is suppressing large excursions along the gauge orbits in the 
imaginary direction. But here it should be noted that ergodicity in the 
complexified configuration space is not necessary for correctness, as can 
be seen already in the simplest one-link U(1) model \cite{arXiv:0912.3360}, where 
the equilibrium configurations contains a delta function in the 
imaginary part. Moreover, ignoring problems of numerical precision, the 
process projected on the space of gauge orbits (the quotient space of 
the full configuration space by the gauge group) is not affected by the 
cooling, so it is as ergodic as the process without cooling.

Finally, we note that the formal justification of the CL method given in Ref.\ 
\cite{arXiv:0912.3360} depends crucially on the holomorphic nature of the 
drift entering the process. Gauge cooling, in the way we implement it, 
certainly cannot be described by such a holomorphic drift. But since we 
need to justify the process only for gauge invariant observables, it is 
only the projection of the process onto the space of gauge orbits that 
matters. The latter, as remarked above, is unaffected by gauge cooling if 
we ignore problems of numerical precision.

\begin{figure}[t]
  \begin{center}
    \includegraphics[width=0.45\textwidth]{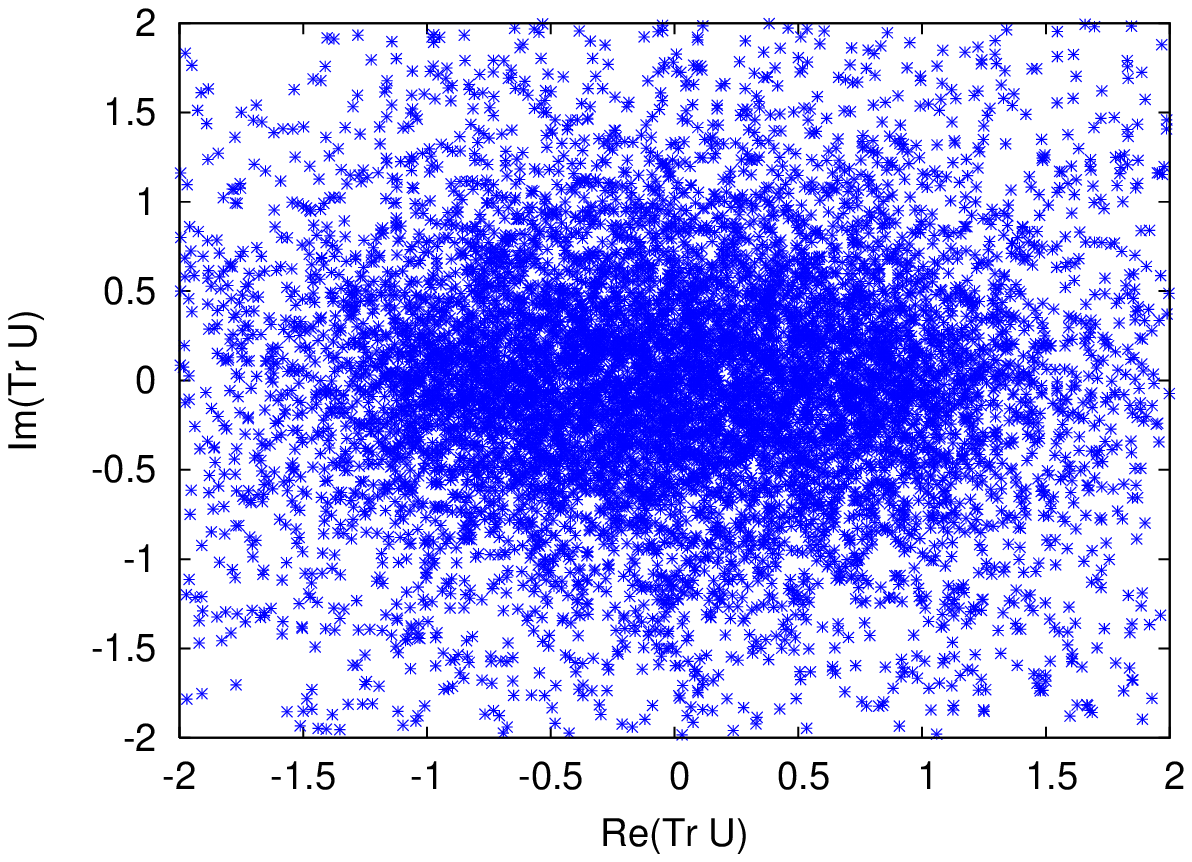} \\
    \includegraphics[width=0.45\textwidth]{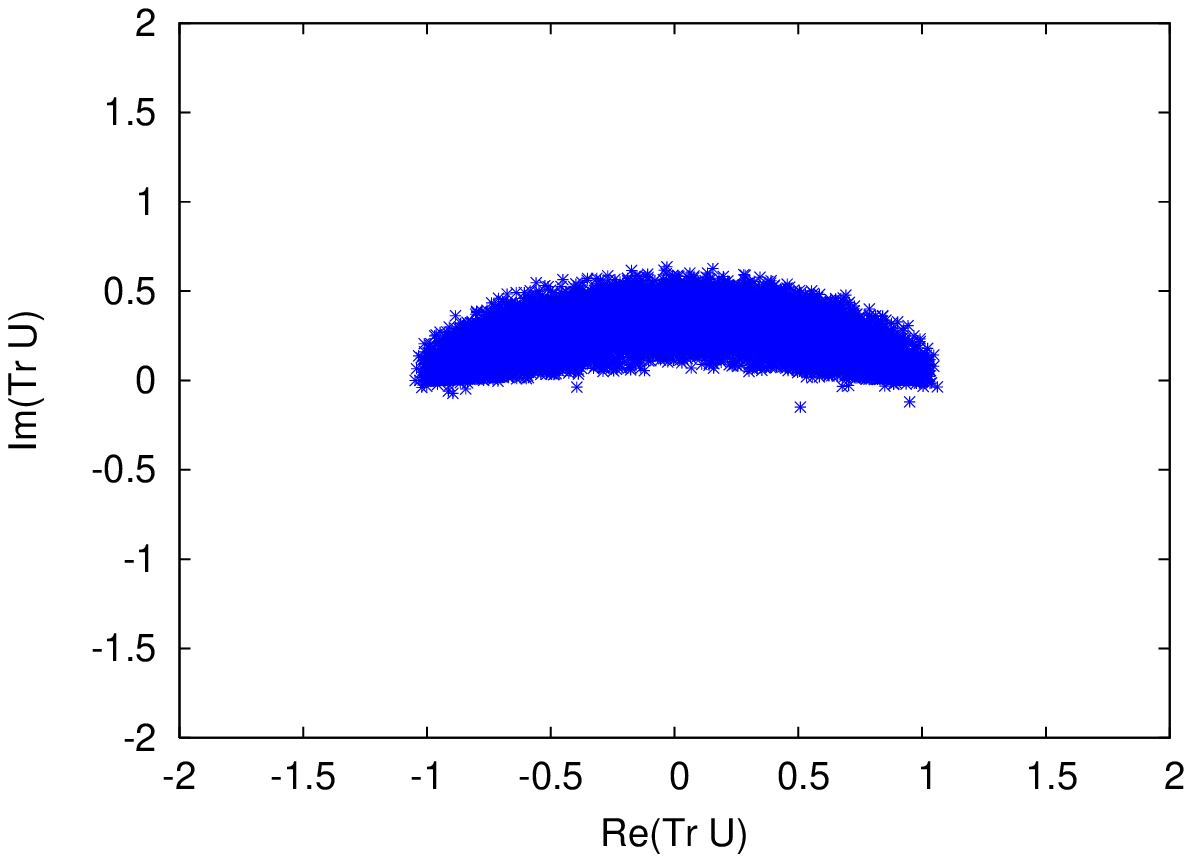}
  \caption{SU(2) one-link model: scatter plots of $\Tr\,U$ without (top) and with (bottom) gauge fixing, at $\beta=i$ \cite{Berges:2007nr}.}
  \label{fig:su2}
 \end{center}
 \end{figure}

As a precursor to gauge cooling, we first discuss a simple implementation to illustrate the effect of gauge transformations on the distribution \cite{Berges:2007nr}. We consider the SU(2) model  (\ref{eq:su2}) with $N_\ell=1$ and complex $\beta=i$, and express the matrix $U$ as
\be
U = c\id +i\sigma_ab_a,
\quad\quad\quad
 c^2+b_ab_a=1,
\ee
where $\sigma_a$ are the Pauli matrices. A CL update reads $U\to U'=RU$, where $R$ was defined in Eq.\ (\ref{eq:R}). A scatter plot of $\Tr\,U$ during the CL evolution is shown in Fig.\ \ref{fig:su2} (top). The distribution is very broad and not well-localised: indeed the resulting expectation value $\bra\Tr\,U\ket$ disagrees with the exact result. However, the model only depends on  $\Tr\,U = 2c$ and the full evolution of $U$ contains therefore considerable gauge freedom. This can be constrained by diagonalising $U$ after each CL update (gauge fixing). The resulting scatter plot, Fig.\ \ref{fig:su2} (bottom), shows a distribution which is extremely well-localised, it is effectively zero outside the banana shape, and the  expectation value $\bra\Tr\,U\ket$ now agrees with the exact result  \cite{Berges:2007nr}. We conclude that a combination of CL updates and gauge transformations has the desired effect.

\begin{figure}[t]
 \begin{center}
 \includegraphics[width=0.39\textwidth]{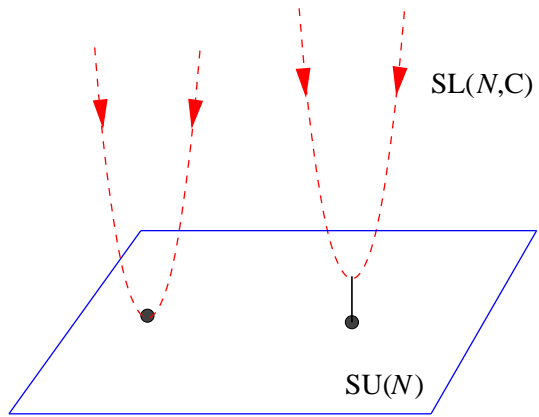} 
 \end{center}
\caption{Gauge cooling in SL($N, \mathbb{C}$) brings the links as close as possible to SU($N$). The orbit on the left is equivalent to a SU($N$) configuration, while the one on the right is not.
}
\label{figslnc}
 \end{figure}

The simple gauge fixing used above does not generalise to  gauge theories in an obvious way. 
In Ref.\ \cite{Berges:2007nr} it was shown that using gauge 
fixed simulations in the maximal axial gauge helps in decreasing the 
width of the link distribution in the imaginary directions. Here, 
however, we use gauge cooling to control the deviation from the unitary 
submanifold without specifying a conventional gauge condition.
This can be achieved by choosing $\om_a$ as the gradient of a suitably chosen function, such as the distance $\dd$ or the unitarity norm. In principle $\om_a$ is complex; however, 
choosing $\omega_a$ (anti)parallel to the gradient
yields  a unitary gauge transformation which is of no interest here and we therefore take $\om_a$ to be purely imaginary. 
Starting from the distance
\be
\label{eq:d}
\dd = \frac{1}{N_\ell}\sum_{k=1}^{N_\ell} \frac{1}{N}\Tr \left(U_kU_k^\dagger - \id\right),
\ee
we arrive at the following cooling update at site $k$, 
\bea
&& U_k\to  U_k' = \Omega_k U_k, 
\nn
\\
&& U_{k-1}\to  U_{k-1}' =U_{k-1}\Omega_k^{-1},
 \eea
 with 
 \be
 \Omega_k = e^{-\eps\alpha\lambda_af_a^k}, 
 \ee
 where
 \be
f_a^k = 2\Tr\, \lambda_a\left(U_kU_k^\dagger - U_{k-1}^\dagger U_{k-1}\right).
\ee
Here $\eps$ is the Langevin time step and $\alpha$ is a (positive) parameter, which can be chosen at will. Below we show that choosing $\alpha$ adaptively can lead to very fast evolution towards the unitary submanifold.

In order to demonstrate that a cooling update indeed brings the configuration closer to SU($N$), we consider a cooling update at site $k$ and find that the distance $\dd$ changes, to first order in $\eps\alpha$, as
\be
\dd '-\dd =  -\frac{\eps\alpha}{NN_\ell} (f_a^k)^2 +{\cal O}\left((\eps\alpha)^2\right). 
\ee
If the original configuration is gauge-equivalent to a unitary one, cooling will eventually transform the configuration into a unitary one. Since $f_a^k=0$ for a  unitary configuration, cooling then no longer has an effect.
If the configuration is not gauge-equivalent to SU($N$), cooling will bring it as close as possible, i.e.\ it will minimize $\dd$. These ideas are sketched in Fig.\ \ref{figslnc}.

Before turning to a numerical solution of CL combined with gauge cooling, we first analyse cooling in  a one-link model analytically. Since we only consider cooling and no drift, the form of the action does not come into play; however, for definiteness one may consider the models in Eqs.\  (\ref{eq:su2}) and (\ref{eq:actsu3}) with $N_\ell=1$. 
After taking the continuous Langevin time limit ($\eps\to 0$), the change in $\dd$, to leading order in $\alpha$, is given by
\be
\dot \dd = -\frac{\alpha}{N} f_a^2, 
\ee 
with
\be
f_a =  2\Tr\, \lambda_a\left(UU^\dagger - U^\dagger U\right).
\ee
Using the Fierz identity,
\be
\lambda_a^{ij}\lambda_a^{kl} = 2\left( \delta_{il}\delta_{jk} - \frac{1}{N}\delta_{ij}\delta_{kl}\right),
\ee
this takes the form
\be
\dot \dd = -\frac{16\alpha}{N} \Tr\, UU^\dagger[U,U^\dagger].
\ee
Specialising to SU(2) to simplify the expression for the trace, we find
\be
\dot \dd =  -8 \alpha \left( \dd^2+2\left(1-|c|^2\right)\dd+c^2+c^{*2}-2|c|^2\right),
\ee
where 
\be
c =\half\Tr\, U, \quad\quad\quad c^* =\half\Tr\, U^\dagger.
\ee
To understand this expression, we first note that in the one-link model $c$ and $c^*$ are invariant under gauge transformations.  Furthermore, for an SU(2) matrix, $c=c^*$. Therefore a matrix $U\in$ SL(2, $\mathbb{C}$) with a complex trace cannot be gauge-equivalent to an SU(2) matrix:  $\dd$ should then remain larger than 0 under cooling. If on the other hand the trace is real, the evolution equation simplifies to
\be
\dot \dd =  -8 \alpha\left(\dd+2-2c^2\right) \dd,
\ee
with the unitary fixed point $\dd=0$ (the other fixed point cannot be reached, since $\dd+2> 2c^2$; here and below we exclude the trivial case of the identity matrix, with $c=c^*=1$). From the explicit solution,
\be
\dd(t) = c^2-1+(1-c^2)\coth\left( 8 \alpha(1-c^2)t+\rm{const.}\right),
\ee
we see that the unitary fixed point is reached exponentially fast,
\be
 \dd(t)  \sim 2\left(1-c^2\right)e^{-16\alpha \left(1-c^2\right)t}\to 0.
\ee
For an SL(2, $\mathbb{C}$) matrix which is not gauge-equivalent to an SU(2) matrix, with $c^*\neq c$,  the fixed point of the cooling evolution is at 
\be
\label{eq:523}
\dd_0 = |c|^2-1+\sqrt{1-c^2-c^{*2}+|c|^4}>0,
\ee
which is again reached exponentially fast.
We conclude that in the SU(2) one-link model, gauge cooling results in an exponential reduction of 
$\dd-\dd_0$, where $\dd_0$ is the minimum given in Eq.\ (\ref{eq:523}).
    We note that an exponential approach is to be expected on general grounds because $\dd$ (as well as the other norm)  is a smooth function of the gauge transformations; hence the gradient will vanish linearly at the minimum on the gauge orbit.

The cooling procedure itself can be improved similarly to the dynamics.
One method is to use an adaptive step size (equivalently, an adaptive 
cooling parameter $\alpha$). We shall discuss this in connection with the numerical results below. 
Another method is Fourier acceleration \cite{Batrouni:1985jn}. 
To demonstrate this, we consider a lattice with $L^D$ sites and
discuss first the case of U(1) with links
\be
U_{x,\mu}= \exp [iA_{x,\mu}+B_{x,\mu}].
\ee
Since the distance as defined above is not bounded by 0 for the U(1) 
case, we use the symmetrised version
\bea
\dd = &&\hm \frac{1}{2}\sum_{x,\mu} \left(U_{x,\mu}^*U_{x,\mu} + 
{U_{x,\mu}^{-1}}^*U_{x,\mu}^{-1} -2\right)
 \nn\\
  =&&\hm \sum_{x,\mu} \left(\cosh[2 B_{x,\mu}] -1\right) \geq 0.
\eea
We consider the gauge transformation
\be
B_{x,\mu} \longrightarrow B_{x,\mu} + y_{x+\mu} - y_x.
\ee
Let the minimum of $\dd$ over gauge transformations be attained for
$B=B^0$ and denote the minimal value of $\dd$ by $\dd_0$. Then, writing
\be
B_{x,\mu}=  B^0_{x,\mu} + y_{x+\mu} - y_x,
\ee
the gradient of $\dd$ with respect to the gauge parameters is 
\bea
\nn
\frac{\partial}{\partial y_x}\dd =  - \sum_{\mu} \big( 
\sinh[2B^0_{x,\mu}+2y_{x+\mu}-2y_x]  &&\\
 -\sinh[2B^0_{x-\mu,\mu}+2y_x-2y_{x-\mu}] \big). &&
\label{grad}
\eea
Because the gradient has to vanish if all $y_x=0$, $B^0$ satisfies 
something like a Landau gauge condition:
\be
\label{landau}
\sum_\mu \left({\sinh[2 B^0_{x,\mu}]- \sinh[2 B^0_{x-\mu,\mu}]} \right) = 0
\ee
for all $x$. Linearising Eqs.\ (\ref{grad}), (\ref{landau}) both in the 
gauge parameters $y_x$ and the fields $B^0_{x,\mu}$ we obtain 
\be
\frac{\partial}{\partial y_x} \dd = -2 \sum_{\mu} \left(y_{x+\mu} -2y_x + 
y_{x-\mu} \right) =-2\Delta y,
\ee
where $\Delta$ is the lattice Laplacean. In this approximation thus 
\bea
\dd = \dd_0 + 2\sum_{x,\mu} (\grad y)^2 = \dd_0 - 2(y,\Delta y).
\eea
Without Fourier acceleration the gauge cooling equation reads, to the 
same order,
\be
\dot \dd = - \alpha (\Delta y,\Delta y),
\ee
which gives an exponential approach to the minimum
with the rate determined by the softest mode $k=2\pi/L$.

To implement Fourier acceleration we calculate the Fourier transform of 
the gradient, divide it by $\hat k^2=\sum_\mu (2-2 \cos k_\mu)$ and 
transform it back using the inverse Fourier transform. This leads to
\be
\dot \dd  =  \alpha (y,\Delta y) = -\alpha\left( \dd-\dd_0\right),
\ee
which shows again an exponential approach, but now with a rate $\alpha$. 
For the nonabelian case we just use this procedure for every colour, 
i.e.\ the variables $B_{x,\mu},y_x$ acquire a colour index~$a$; to  
leading order in those variables the method is justified just as for U(1).

 \begin{figure}[t]
 \begin{center}
 \includegraphics[width=0.45\textwidth]{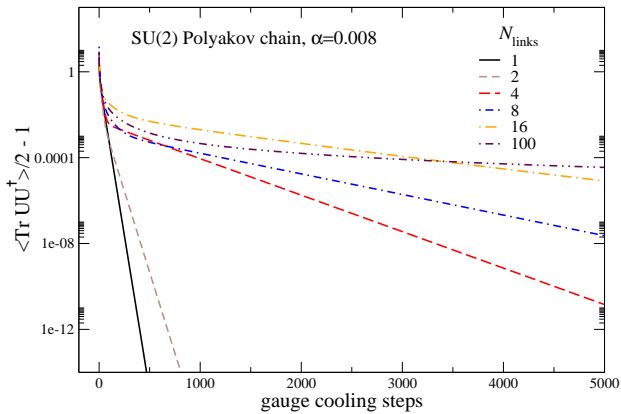} 
 \end{center}
\caption{SU(2) Polyakov chain: evolution of $\dd$ without CL updates, at a fixed gauge cooling parameter $\alpha=0.008$, for various values of $N_\ell$, starting from a gauge transformed unitary chain.
}
\label{fig:su2N}
 \end{figure}

\subsection{Gauge cooling for the SU(2) Polyakov chain}
\label{sec:coolsu2}

 We now turn to the numerical solution of the SU(2) Pol\-ya\-kov chain (\ref{eq:su2}) with $N_\ell$ links. We start by considering  cooling only, i.e.\ there is no actual CL step in the update. Since only the combination $\eps\alpha$ appears in the cooling update, we take $\eps=1$. In Fig.\ \ref{fig:su2N}, we show the dependence of the distance $\dd$, see Eq.\ (\ref{eq:d}), on the number of cooling steps, for a number of chains with $N_\ell$ links.
 As initial condition we use an SL(2, $\mathbb{C}$) gauge transformed unitary chain, i.e.\
 \be 
 U_k(0) = \Omega_kV_k\Omega_{k+1}^{-1}, 
 \quad\quad\quad
 V_k\in \mbox{SU($2$)},
\ee
with $\Omega_k\in\mbox{SL(2, $\mathbb{C}$)}$  a random gauge transformation.
 Cooling is therefore expected to return the chain to being unitary, with $\dd=0$. 
 This is indeed visible in Fig.\ \ref{fig:su2N} and for a small number of links this happens exponentially fast (note the vertical log scale).
 The rate, however, decreases as the number of links is increased and eventually, for $N_\ell=100$, the reduction appears to become nonexponential.

\begin{figure}[t]
 \begin{center}
 \includegraphics[width=0.45\textwidth]{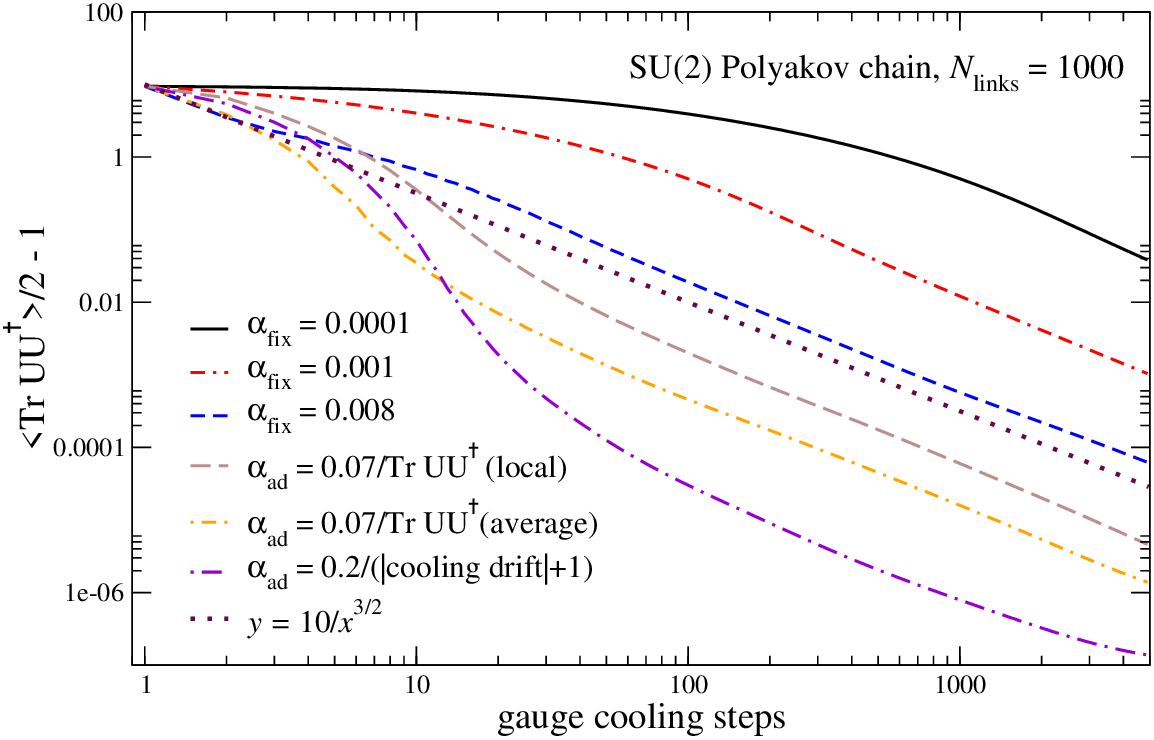} 
 \end{center}
\caption{
As in Fig.\ \ref{fig:su2N},  for various values of the gauge cooling parameter $\alpha$ and various adaptive implementations, with $N_\ell=1000$, on a log-log scale.
The dotted  line indicates a power decay, with power $3/2$.
}
\label{fig:su2c2}
 \begin{center}
 \includegraphics[width=0.45\textwidth]{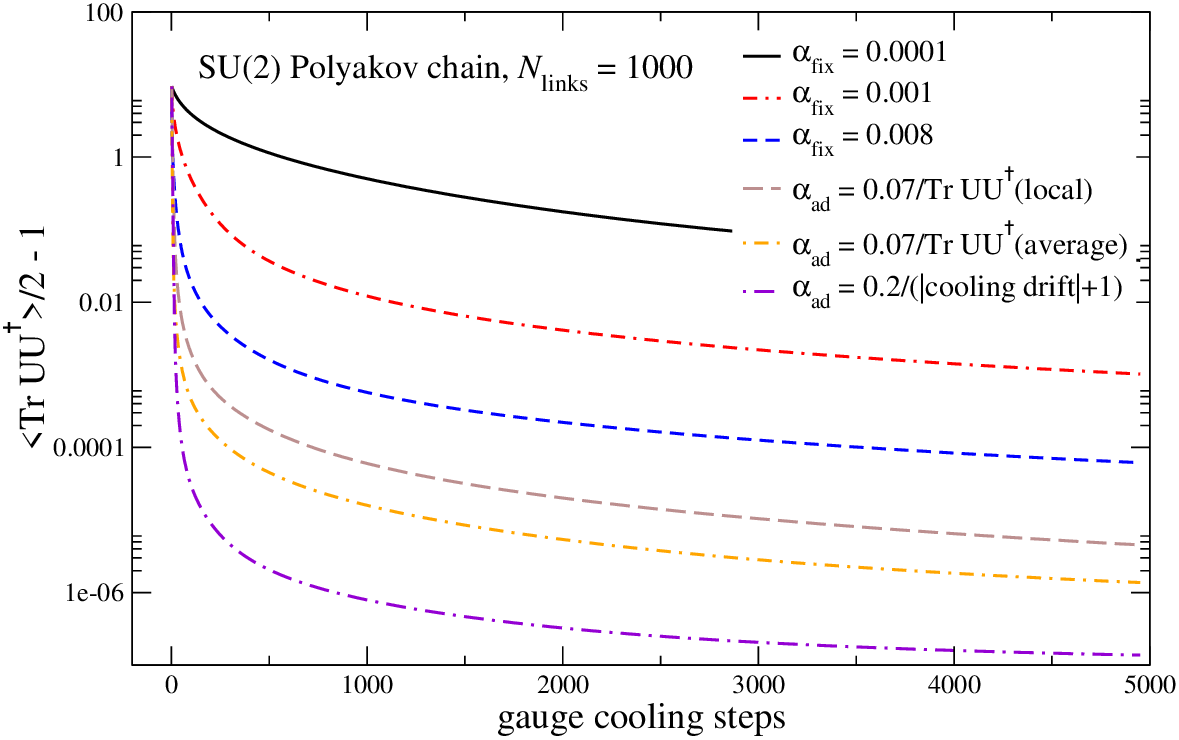} 
 \end{center}
\caption{As in Fig.\ \ref{fig:su2c2}, on a linear-log scale.
}
\label{fig:su2c1}
 \end{figure}

This is further demonstrated in Fig.\ \ref{fig:su2c2}, where the evolution under cooling is shown for a long chain with $N_\ell=1000$ links, starting again from a gauge transformed unitary chain.
The top three lines are obtained as above, using three values of $\alpha$: a larger $\alpha$ results in a faster cooling, as expected. 
However, in all cases we note that the decay is power-like rather than exponential, with a power $3/2$, as indicated with the dotted line (note the log-log scale).
It is therefore useful to optimise the reduction of $\dd$, after a given number of cooling steps, by employing an adaptive cooling algorithm. Here the guiding principle is to use an effective adaptive $\alpha_{\rm ad}$ which is as large as possible. However, choosing $\alpha_{\rm ad}$ too large may lead to instabilities when the cooling drift is large as well, i.e.\ far away from the unitary submanifold. We therefore control $\alpha_{\rm ad}$ by moderating it appropriately. Explicitly,
we have implemented adaptive schemes of the form
\be
\alpha_{\rm ad} =  \frac{\alpha}{D(U, U^\dagger)},
\ee
where $D(U, U^\dagger)$ depends on the links at the present cooling step. In Fig.\ \ref{fig:su2c2} we present results for $D=\Tr\,UU^\dagger$, both locally at site $k$ as well as averaged over the chain, and  $D$ determined by the absolute value of the cooling drift, $D=\sum_{ak}|f_a^k|/N_\ell +D_0$, with $D_0$ a constant.
When the chain is close to being unitary, both choices reduce to a constant effective $\alpha$, with $\alpha_{\rm ad} = \alpha$ and $\alpha/D_0$, respectively.
From Fig.\ \ref{fig:su2c2} we note that the adaptive cooling results also contain an interval with sub-exponential decay,  however they are much more effective at the earlier stages of cooling. This is demonstrated in Fig.\ \ref{fig:su2c1}, where the same data is presented, but now using a log-linear scale. With an adaptive implementation, after only a few cooling steps the distance from SU(2) is greatly reduced, drastically changing the distribution in the complexified space, as we will see now.

   \begin{figure}[t]
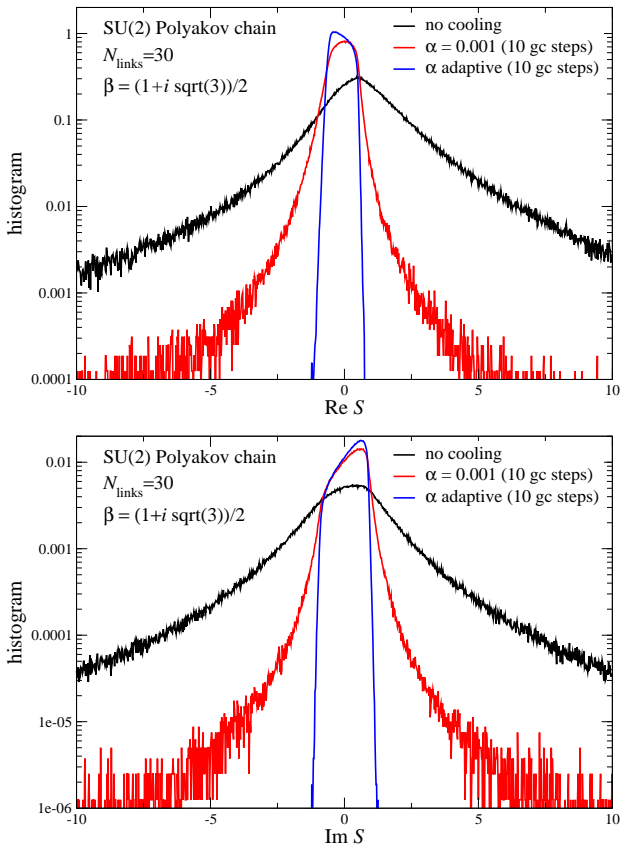

 \begin{center}
 \includegraphics[width=0.45\textwidth]{ReS_Bin10000.eps} 
 
 \vspace*{0.2cm}
 
  \includegraphics[width=0.45\textwidth]{ImS_Bin10000.eps} 
 \end{center}
\caption{SU(2) Polyakov chain: histogram of the real (top) and imaginary (bottom) part of the action, obtained during a CL simulation using no cooling (broadest distribution), cooling with a constant $\alpha$ (middle) and cooling with an adaptive $\alpha_{\rm ad}=0.2/(|$cooling drift$|+0.8)$ (narrow distribution), using $N_\ell=30$ links at $\beta=\half(1+i\sqrt{3})$. Note that the value of $\alpha$ represents the product $\eps\alpha$.  
}
\label{fig:su2hist}
 \end{figure}

To proceed, we now include CL evolution with a complex drift and consider an SU(2) chain with $N_\ell=30$ links at $\beta=\half(1+i\sqrt{3})$. There is a sign problem due to the complex coupling.
In order to control the distribution, each CL update is followed by a number of gauge cooling steps. In Fig.\ \ref{fig:su2hist} we present distributions for the real and imaginary parts of the action $S$, where the data is collected after performing a CL update followed by a series of gauge cooling updates. Three histograms are shown: one without cooling, one with ten cooling steps at a fixed $\alpha=0.001$ and one with  ten cooling steps using an adaptive $\alpha_{\rm ad}=0.2/(|$cooling drift$|+0.8)$, where the cooling drift is averaged over the chain (note that the value of $\alpha$ represents the product $\eps\alpha$).  
We see significant changes in the distribution: without cooling the distribution is wide and slowly decreasing; with fixed $\alpha$ the distribution is already more localised but skirts are still present. Finally with the adaptive choice, the distribution is localised very well and drops quickly to zero outside its main support. 

\begin{figure}[t]
 \begin{center}
 \includegraphics[width=0.45\textwidth]{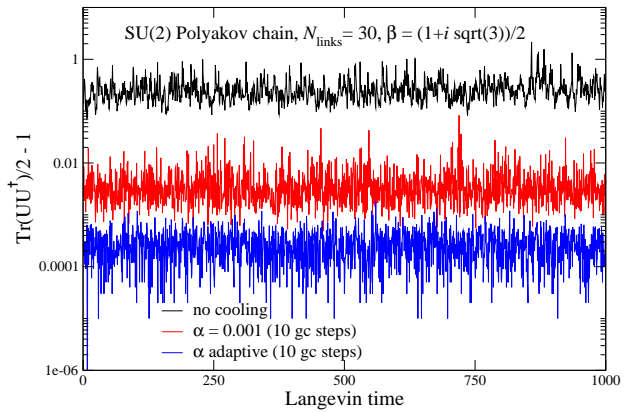} 
 \end{center}
\caption{SU(2) Polyakov chain: Langevin time evolution of $\dd$, without cooling  and using the cooling algorithms of Fig.\ \ref{fig:su2hist}.
}
\label{fig:su2evol}
 \end{figure}

 These features are also reflected in the distance $\dd$ from the unitary submanifold. This is demonstrated in Fig.\ \ref{fig:su2evol}, where we show the Langevin time evolution of $\dd$.   Without cooling, we find that the average distance equals $\bra\dd\ket=0.26(2)$, i.e.\  relatively far away from the unitary submanifold. Employing ten steps of cooling with a constant $\alpha=0.001$ reduces this to $\bra\dd\ket= 0.00425(2)$. Finally with the adaptive cooling implementation this is reduced even further to $\bra\dd\ket\sim 0.00029822(4)$. Cooling therefore brings the configuration closer to SU(2).
Since the coupling and therefore the action is complex, it should be noted that $\dd$ necessarily has to exceed 0, but apparently it can be very close to 0.

\begin{figure}[t]
 \begin{center} 
 \includegraphics[width=0.42\textwidth]{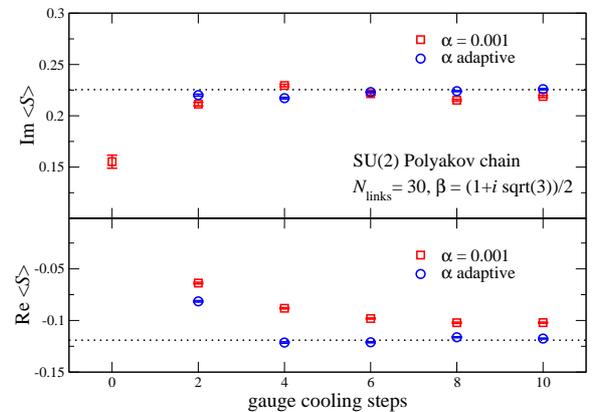} 
 \end{center}
\caption{SU(2) Polyakov chain: imaginary (top) and real (bottom) part of the expectation value of the action as a function of the number of cooling steps used between CL updates, for the two cooling algorithms of Fig.\ \ref{fig:su2hist}. The dotted lines indicate the exact results.
}
\label{fig:su2S}
 \end{figure}

Finally, in order to see whether the CL simulation yields a sensible result, we compare the expectation value of the action with the exact result  at $\beta=\half(1+i\sqrt{3})$, obtained by direct numerical integration,
\be
\bra S\ket_{\rm exact} = -0.119121+i0.225487.
\ee
The result is shown in Fig.\ \ref{fig:su2S} as a function of the number of cooling steps used between the CL updates. We note that the data represent results from simulations using a number of Langevin stepsizes and performing an extrapolation to zero stepsize.
For the real part, the result without cooling is far off the exact result and not shown. With fixed $\alpha=0.001$ we observe that the outcome improves  when the number of cooling steps is increased but there is still disagreement after using ten gauge cooling steps. On the other hand, using the adaptive implementation, we find agreement with the exact results, provided that at least six cooling steps are used between CL updates.
This agreement supports our insight based on the requirement of the localisation of distributions in the complexified space.

\begin{figure}[t]
  \includegraphics[width= 1.\columnwidth]{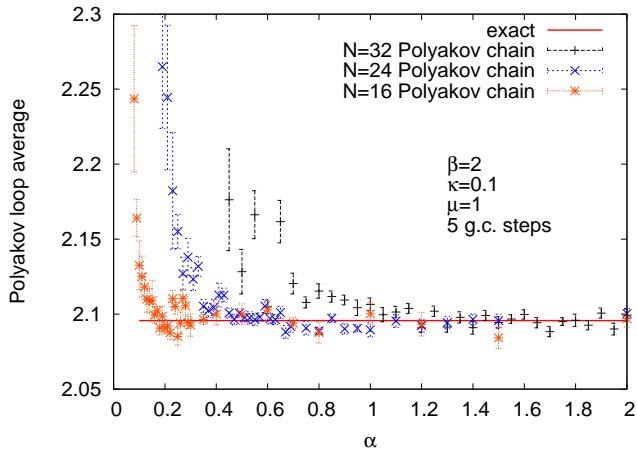}
  \caption{SU(3) Polyakov chain: Polyakov loop expectation value as a function of gauge cooling parameter $\alpha$, using 5 gauge cooling steps, for various values of $N_\ell$, at fixed $\beta, \kappa$ and $\mu$ \cite{Seiler:2012wz}. }
\label{fig:su3polav}
\end{figure}

\subsection{Gauge cooling for the SU(3) Polyakov chain}
\label{sec:coolsu3}

Since the SU(3) theory at nonzero chemical potential represents a genuinely complex problem, we need to verify that the understanding gained from the SU(2) case also applies to the SU(3) case. We first consider the Polyakov chain model with the action (\ref{eq:actsu3}). We refer to
\be
 P = \frac{1}{3}\Tr\left( U_1\ldots U_{N_\ell}\right)
\ee
as the Polyakov loop. We use again as many as 1024 links to simulate the number of variables of a small lattice (the results as shown on the smaller chains are, however, already representative).
Since the procedure is similar to the SU(2) case we here concentrate
on some results. More details can be found in Ref.\  \cite{Seiler:2012wz}.

\begin{figure}[t]
\begin{center}
\includegraphics[width=0.9\columnwidth]{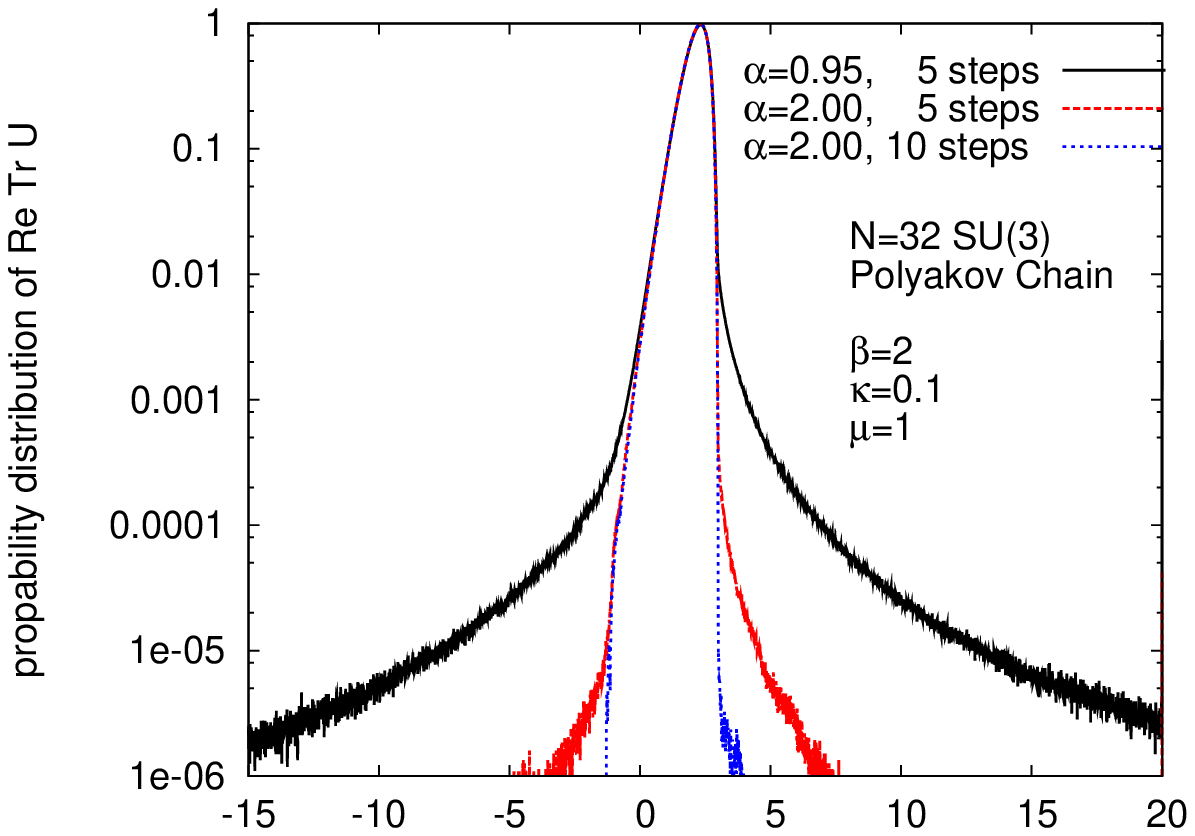}  
\caption{SU(3) Polyakov chain: histograms of the real part of the Polyakov loop, using
various combinations of $\alpha$ and the number of cooling steps, with $N_\ell=32$ links \cite{Seiler:2012wz}.}
\label{fig:su3polhist}
\end{center}
\begin{center}
\includegraphics[width=0.98\columnwidth]{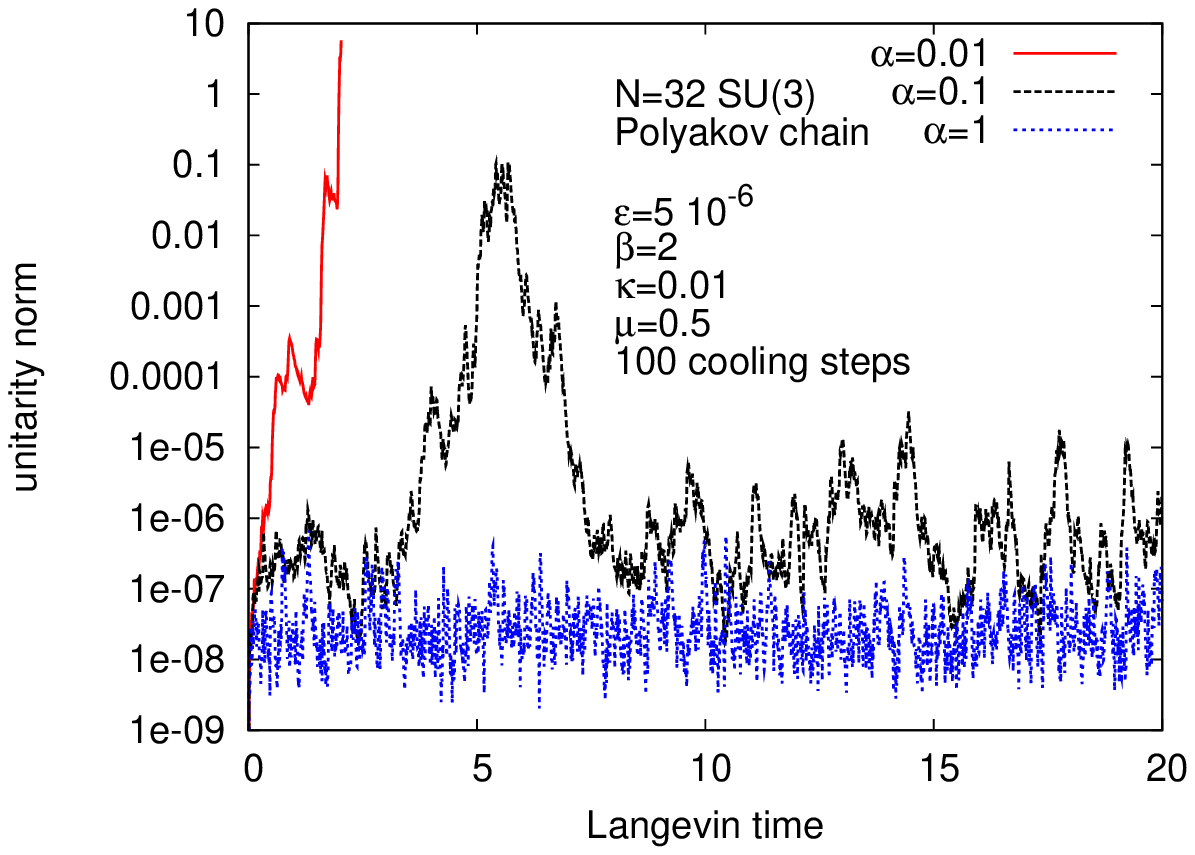}  
\caption{SU(3) Polyakov chain: Langevin time evolution of the unitarity norm, using various $\alpha$ values ($N_\ell=32$).}
\label{fig:su3unhist}
\end{center}
\end{figure}

In Fig.\ \ref{fig:su3polav} we compare the expectation value of the Polyakov loop $\bra P\ket$
  with the exact result, for three chains with 16, 24, and 32 links. Between each CL update 5 gauge cooling steps are inserted, with different choices for the gauge cooling parameter $\alpha$ to enhance the efficiency of the cooling. As above, the exact result is reproduced when sufficient cooling is applied.
As shown in Fig.\ \ref{fig:su3polhist}, this conclusion is consistent with results for the Polyakov loop distribution for the various $\alpha$'s. Cooling drastically reduces the ``skirt" of the distribution to a level at which it becomes harmless, as corroborated by the results in Fig.~\ref{fig:su3polav}.

We have used both the distance (\ref{eq:d1}) and the unitarity norm (\ref{eq:norm}) as the function to be minimised under cooling.
In Fig.\ \ref{fig:su3unhist} the Langevin time evolution of the unitarity norm, averaged over the chain, is shown. Here we applied 100 cooling steps between each Langevin update, using again various values of $\alpha$.
The cooling dynamics is determined by the gradient of the unitarity norm.
 As above we find that substantial cooling stabilises the dynamics and keeps the distance from the unitary submanifold under control.

 \begin{figure}[t]
\begin{center}
\includegraphics[width=0.9\columnwidth]{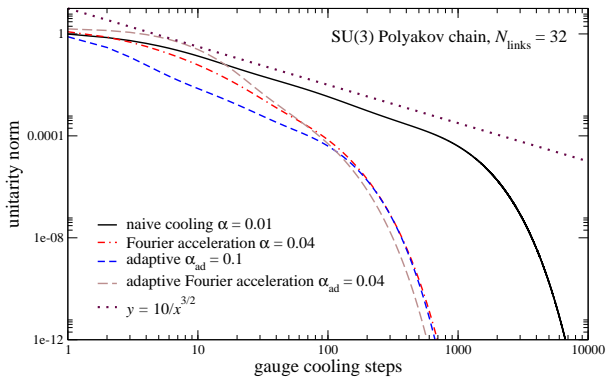}  
\caption{SU(3) Polyakov chain: evolution of the unitary norm without CL updates, starting from a gauge transformed unitary chain, using various cooling implementations ($N_\ell=32$).}
\label{fig:su3coolev}
\end{center}
\end{figure}

As in the SU(2) case, the evolution of cooling typically shows two regimes, a power law behaviour at earlier times followed by an exponential falloff. We observe the same power $3/2$ as in the SU(2) Polyakov chain.
With a nonadaptive $\alpha$, the power law regime dominates for a long time, see Fig.\ \ref{fig:su3coolev} (the length of the power law regime increases rapidly with the number of links).
The efficiency of cooling can be strongly enhanced by observing that the speed with which the minimum is approached is determined by the smallest nonzero Fourier mode.
Adaptive cooling and Fourier acceleration 
as described above can strongly increase the efficiency of cooling, as demonstrated in Fig.\ \ref{fig:su3coolev}.
We conclude therefore that the gauge cooling is as efficient as in
 the SU(2) case and allows one to reach the correct results.

\subsection{Gauge cooling for QCD with heavy quarks}
\label{sec:coolhdm}

 As our final step towards full QCD, we consider QCD at nonzero chemical potential with heavy quarks (HQCD). This theory retains many of the features of QCD and represents an approximation
 to the latter in the large mass, large chemical potential region (heavy dense QCD) \cite{bender,feo}. For Wilson fermions, with hopping parameter $\kappa$, it is based on a resummed hopping parameter expansion,
retaining only time-like Polyakov loops, which can formally be obtained from the 
 limit $\kappa \rightarrow 0$, $\mu \rightarrow \infty$, with  $\kappa\e^{\mu}$ fixed.
 This approximation, and variants thereof, have repeatedly been used to obtain
 information about the phase structure of QCD in the $\mu - T$
 plane \cite{feo,Fromm:2011qi}. Here we use a 
``symmetrized" version  \cite{Aarts:2008rr} of the model, with the action
  \be
S = S_{\rm YM} - \ln \det {\bf M}(\mu).
\ee 
Here $S_{\rm YM}$ is the standard Wilson action on a $N_s^3\times N_\tau$ lattice with gauge coupling $\beta$, and the determinant takes the form
\be
\det {\bf M} =  \prod_\vecx 
\det \left(1 + h e^{\mu/T} {\cP}_\xv\right)^2 
\det \left(1+ h e^{-\mu/T} {\cP_\xv^{-1}}\right)^2,
\ee
where $T=1/N_\tau$, lattice units are used throughout, $h=(2\kappa)^{N_\tau}$, and the (conjugate) Polyakov loops are
\be 
\cP_\xv= \prod_{\tau=0} ^{N_\tau-1}U_{(\tau,\xv),4},
\quad\quad\quad
\cP_\xv^{-1} \cP_{\xv}  = \id.
\ee
Below we also consider the traced Polyakov loops, $P_\xv^{(-1)} = \frac{1}{3}\tr\, \cP_\xv^{(-1)}$.
The factor containing $e^{-\mu/T}$ is of course negligible in the heavy dense limit, but it is required for the symmetry  $[\det {\bf M}(\mu)]^*= \det {\bf M}(-\mu^*)$.

\begin{figure}[t]
  \includegraphics[width= 0.88\columnwidth]{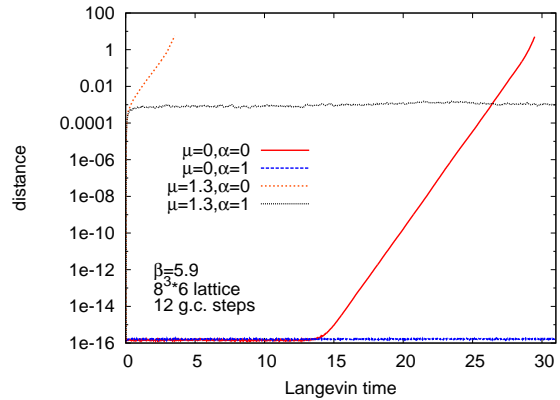}
  \caption{QCD with heavy quarks  (HQCD): evolution of the distance $\dd$ with and without gauge cooling,  at $\mu=0$ and 1.3, for  $\beta=5.9$ on a $8^3\times 6$ lattice, $\kappa=0.12$ throughout \cite{Seiler:2012wz}.}
\label{fig:hdqcdun}
\end{figure}

A detailed analysis can be found in Ref.\  \cite{Seiler:2012wz}: here we discuss a selection of results, some of which are new.
As shown in Fig.\ \ref{fig:hdqcdun}, the effect of gauge cooling  is as important as in the Polyakov chain model. In this figure we present the Langevin time evolution of the (symmetrised version of the) distance \cite{Seiler:2012wz}, for a number of different cases. When $\alpha=0$, no cooling is applied. The unitary submanifold is unstable and the distance increases exponentially. At nonzero $\mu$ this happens immediately, while at vanishing $\mu$  it takes some time for the instability to set in. At vanishing $\mu$ this growth can of course easily be avoided by an occasional re-unitarisation of the links; however, this possibility is not present at nonzero $\mu$ and therefore does not provide a valid solution. With cooling ($\alpha>0$; we used 12 gauge cooling steps between each CL update), we observe that the distance remains under control and in the case of $\mu=0$, the dynamics remains on the unitary submanifold (within numerical precision).

\begin{figure}
  \includegraphics[width= 0.88\columnwidth]{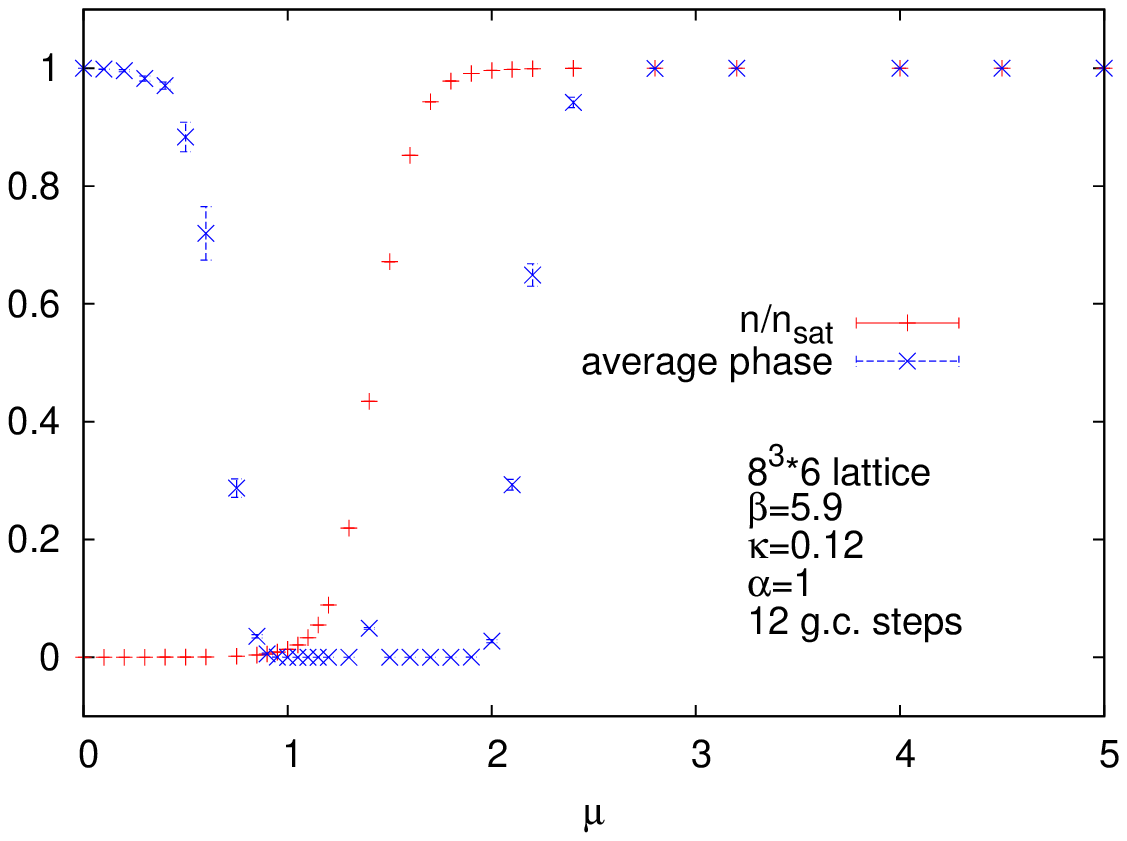}  
  \caption{HQCD: baryon density and average phase factor as a function of $\mu$ for $\beta=5.9$ on  a
   $8^3\times 6$ lattice \cite{Seiler:2012wz}.}
\label{fig:signdens}
  \includegraphics[width= 0.88\columnwidth]{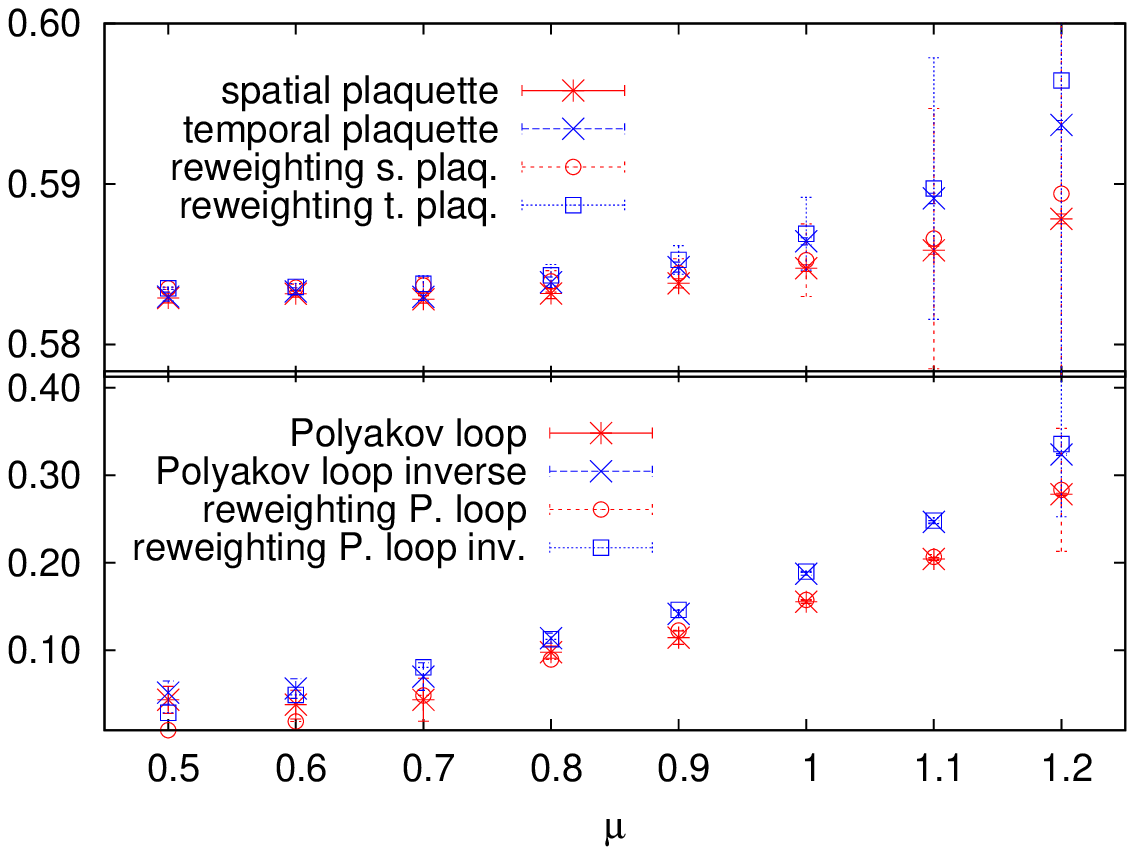}
  \caption{
  HQCD: spatial and temporal  plaquettes (top) and Polyakov loops $\bra P\ket$ and $\bra P^{-1}\ket$ (bottom)
  as a function of $\mu$: comparison between CL and reweighting at 
  $\beta=5.9$ on a $6^4$ lattice, using $\alpha=1$ with 12 
   gauge cooling steps. Large errors at large $\mu$  affect RW only \cite{Seiler:2012wz}.}
\label{fig:rewcheck6}   
\end{figure}

The baryon density, normalized with the density at saturation ($n_{\rm sat}=6$ in this model),  is shown in Fig.\ \ref{fig:signdens} as a function of $\mu$. We observe an onset, followed by a rapid rise to saturation.
We note that the entire region from $\mu=0$ to far into the
 saturation regime is covered. 
Also shown is the average phase factor in the full theory, defined by  
\be
 \left\bra e^{2i\phi} \right\ket = \left\bra\frac{\det {\bf M}(\mu)}{\det{{\bf M}(-\mu)}}\right\ket.
\ee
This observable shows that between onset and saturation the sign problem is severe, such that other  methods such as reweighting will be ineffective here. 
Nevertheless, CL  appears to work well.
We note that the average phase factor is a gauge invariant observable, just as the density or the plaquettes, presented below. The role of cooling in improving results for gauge invariant observables has been discussed at some length below Eq.\ (\ref{eq:gt}).
We note that without (sufficient) cooling, observables do not converge to the expected values, similar as in Figs.\ \ref{fig:su2S} and \ref{fig:su3polav} for the Polyakov chain models.

For QCD with heavy quarks, no exact results or results obtained with sign-problem free approaches are available. We therefore
compare expectation values of (traced) Polyakov loops and temporal and spatial  plaquettes with
a refined reweighting (RW) analysis as described in Ref.~\cite{feo}. 
In Fig.\ \ref{fig:rewcheck6} we show the spatial and temporal plaquettes as well as the (conjugate) Polyakov loops.
Very good agreement is found in the region where reweighting is feasible; for $\mu > 1$ the 
RW data  rapidly deteriorate, due to the sign problem.

\begin{figure}[t]
  \includegraphics[width= 0.85\columnwidth]{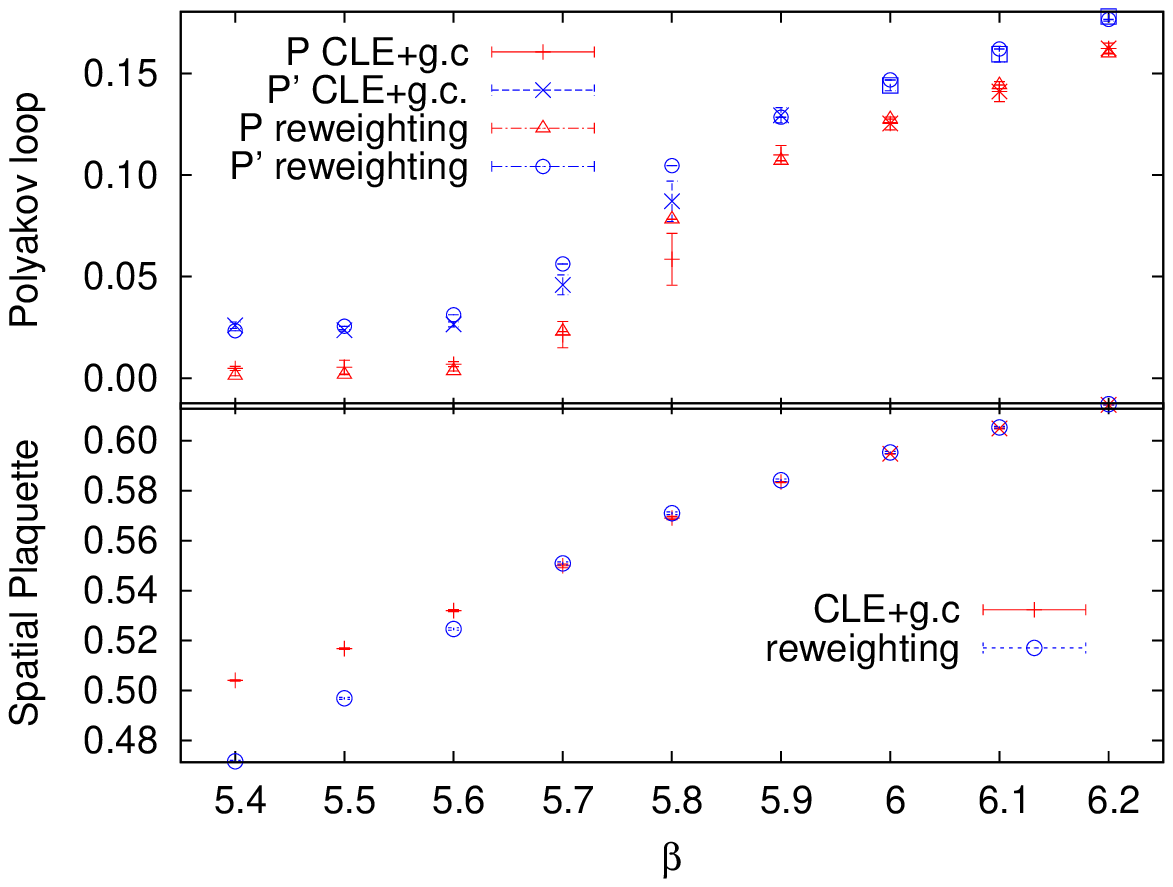}
  \caption{HQCD: Polyakov loops $\bra P\ket$ and $\bra P^{-1}\ket$ (top) and spatial plaquettes (bottom) 
  as a function of $\beta$: comparison between CL and reweighting at 
  $\mu=0.85$ on a $6^4$ lattice \cite{Seiler:2012wz}.}
\label{fig:betasweep}
\vspace*{0.2cm}
  \includegraphics[width= 0.85\columnwidth]{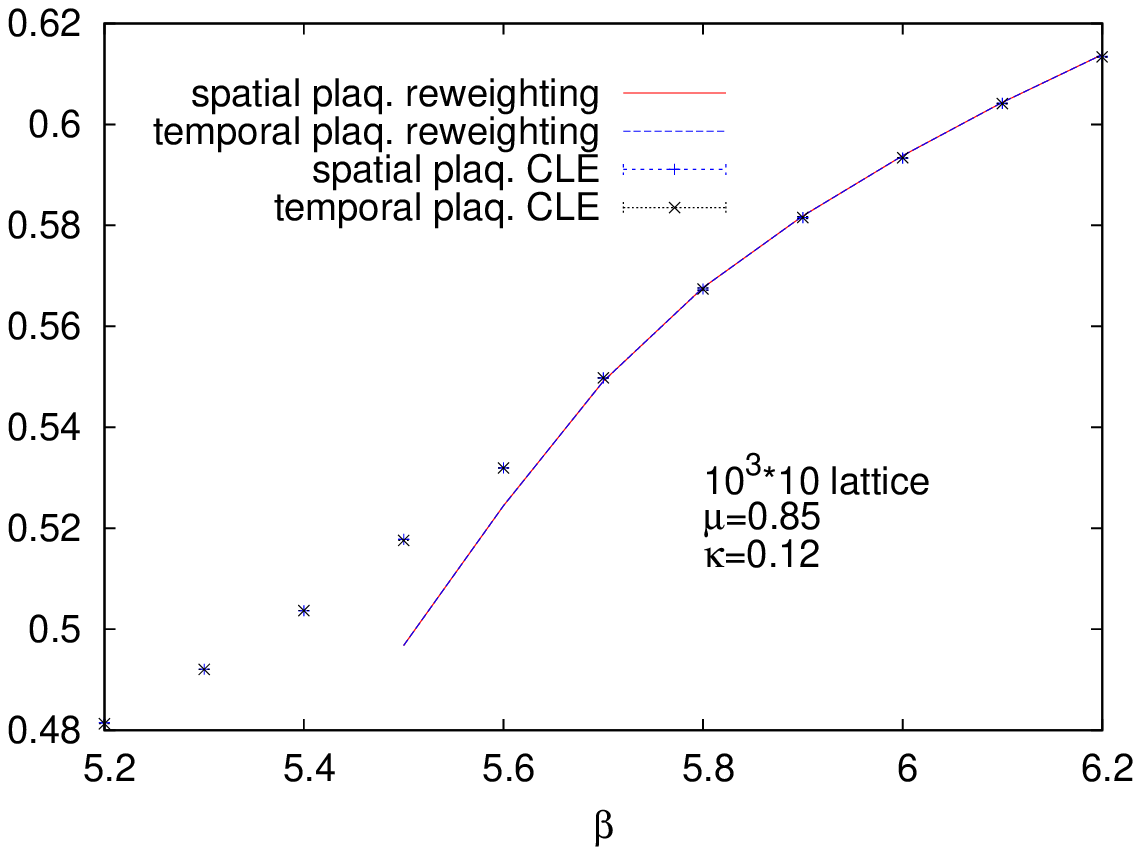}
  \caption{HQCD: temporal and spatial plaquettes as a function of $\beta$:
 comparison between CL and reweighting at  $\mu=0.85$ on a $10^4$ lattice.}
\label{fig:rewcheck10}   
\end{figure}

In Fig.\ \ref{fig:betasweep} we demonstrate that it is possible with CL to go from the deconfined to the confined phase by varying $\beta$ at fixed $\mu$.
 We observe that the Polyakov loop averages agree with the RW results for all $\beta$ values considered.
 For the plaquette expectation values we note that they agree 
 except  for smaller $\beta$ values. In Fig.\ \ref{fig:rewcheck10} we show results on a larger $10^4$ lattice: increasing the lattice size 
 does not appear to move the threshold in $\beta$ below which the plaquette results deteriorate.
First indications are that on the $10^4$ lattice the transition to the confinement region takes place at $\beta \sim 6$.
This suggests that if a deterioration of the CL results occurs, it is a  small (fixed) $\beta$ effect and 
it is possible to go deep into the confining region by increasing the lattice size, a situation very different from the one in 
the XY model.
If this conclusion remains true in full QCD  it means
that we can safely approach the continuum limit of the latter 
in both the confinement and the deconfinement phases, which is an exciting prospect.

\section{Summary}
\label{sec:sum}

Complex Langevin dynamics can solve the numerical sign problem appearing in lattice field theories with a complex action or Boltzmann weight. However, success is not guaranteed. In this paper we reviewed the reasons why a simple proof, valid in the case of real Langevin dynamics, is not available and outlined an alternative route. We indicated how problems can manifest themselves in numerical simulations. Particular emphasis is put on the properties of the positive probability distribution in the complexified space, which is effectively sampled during a CL process. We argued that a proper understanding of this distribution can be used to provide support for the validity of CL results -- or discard the results as a failure.
These insights are demonstrated by contrasting results obtained in two three-dimensional models with  a nonzero chemical potential: the XY model and the SU(3) spin model.

In the second part we applied the lessons learnt to  theories with  SU($N$) gauge symmetry. It is  demonstrated how the freedom to perform SL($N, \mathbb{C}$) gauge transformations can be used in a constructive manner to control the distribution in the extended space.
By interspersing CL updates with SL($N, \mathbb{C}$) gauge cooling steps, the distance from the unitary submanifold can be reduced. We have demonstrated that this can cure the problem of convergence to the wrong result. Adaptive cooling proves to be useful in that it can lead to a surprisingly quick reduction of the deviation from the unitary submanifold, thereby greatly improving the convergence properties of CL.

The gauge cooling techniques discussed here build on the results presented in Ref.\  \cite{Seiler:2012wz}, in which they were applied to QCD with heavy quarks. We are currently extending the analysis, taking into consideration the insights gained here, and hope to report on  our findings in QCD in the near future.

\begin{acknowledgement}
We thank Simon Hands for the opportunity to present this work. 
GA and LB are supported by STFC.
ES and IOS were partially supported by Deutsche Forschungsgemeinschaft.

\end{acknowledgement}


\begin{thebibliography}{}

\bibitem{deForcrand:2010ys}
  P.~de Forcrand,
  PoS LAT {\bf 2009} (2009) 010
  [arXiv:1005.0539 [hep-lat]].

\bibitem{Aarts:2013bla}
  G.~Aarts,
  PoS LATTICE {\bf 2012} (2012) 017
  [arXiv:1302.3028 [hep-lat]].


\bibitem{Parisi:1984cs}
  G.~Parisi,
  Phys.\ Lett.\  B {\bf 131} (1983) 393.

\bibitem{Klauder:1983}
 J.~R.~Klauder,
 Stochastic quantization,
 in: H.~Mitter, C.B.~Lang (Eds.), Recent Developments in High-Energy
Physics, Springer-Verlag, Wien, 1983, p.\ 351;
 J.~Phys.~A: Math.~Gen.~{\bf 16} (1983) L317;
  Phys.~Rev.~A {\bf 29} (1984) 2036.

 
\bibitem{Aarts:2008rr}
  G.~Aarts and I.-O.~Stamatescu,
  JHEP {\bf 0809} (2008) 018
  [arXiv:0807.1597 [hep-lat]].
  
\bibitem{Aarts:2008wh}
  G.~Aarts,
  Phys.\ Rev.\ Lett.\  {\bf 102} (2009) 131601
  [arXiv:0810.2089 [hep-lat]].

\bibitem{arXiv:1006.0332}
  G.~Aarts and K.~Splittorff,
  JHEP\ {\bf 1008} (2010) 017
  [arXiv:1006.0332 [hep-lat]].
  
\bibitem{Aarts:2011zn}
  G.~Aarts and F.~A.~James,
  JHEP {\bf 1201} (2012) 118
  [arXiv:1112.4655 [hep-lat]].
 
 
\bibitem{Ambjorn:1985iw}
  J.~Ambjorn and S.~K.~Yang,
  Phys.\ Lett.\  B {\bf 165} (1985) 140.

\bibitem{Ambjorn:1986fz}
  J.~Ambjorn, M.~Flensburg and C.~Peterson,
  Nucl.\ Phys.\  B {\bf 275} (1986) 375.

\bibitem{Berges:2006xc}
  J.~Berges, S.~Borsanyi, D.~Sexty and I.~O.~Stamatescu,
  Phys.\ Rev.\  D {\bf 75} (2007) 045007
  [hep-lat/0609058].

\bibitem{Berges:2007nr}
  J.~Berges and D.~Sexty,
  Nucl.\ Phys.\  B {\bf 799} (2008) 306
  [arXiv:0708.0779 [hep-lat]].

\bibitem{arXiv:0912.3360}
  G.~Aarts, E.~Seiler and I.~-O.~Stamatescu,
  Phys.\ Rev.\ D\ {\bf 81} (2010) 054508
  [arXiv:0912.3360 [hep-lat]].

\bibitem{arXiv:1005.3468}
  G.~Aarts and F.~A.~James,
  JHEP\ {\bf 1008} (2010) 020
  [arXiv:1005.3468 [hep-lat]].
  
\bibitem{arXiv:1101.3270}
  G.~Aarts, F.~A.~James, E.~Seiler and I.~-O.~Stamatescu,
  Eur.\ Phys.\ J.\ C\ {\bf 71} (2011) 1756
  [arXiv:1101.3270 [hep-lat]].

\bibitem{Pawlowski:2013pje}
  J.~M.~Pawlowski and C.~Zielinski,
  arXiv:1302.1622 [hep-lat];
  arXiv:1302.2249 [hep-lat].

\bibitem{Seiler:2012wz}
  E.~Seiler, D.~Sexty and I.~-O.~Stamatescu,
  arXiv:1211.3709 [hep-lat].

\bibitem{Damgaard:1987rr}
  P.~H.~Damgaard and H.~Huffel,
  Phys.\ Rept.\  {\bf 152} (1987) 227.
  
  \bibitem{reed-simon} 
  M.~Reed and B.~Simon, 
  Methods of Modern Mathematical  Physics IV: Analysis of Operators, Academic Press, New York 1978.

\bibitem{Aarts:2009hn}
  G.~Aarts,
  JHEP {\bf 0905} (2009) 052
  [arXiv:0902.4686 [hep-lat]].

\bibitem{Duncan:2012tc}
  A.~Duncan and M.~Niedermaier,
  arXiv:1205.0307 [quant-ph].

\bibitem{arXiv:0912.0617}
  G.~Aarts, F.~A.~James, E.~Seiler and I.~-O.~Stamatescu,
  Phys.\ Lett.\ B\ {\bf 687} (2010) 154
  [arXiv:0912.0617 [hep-lat]].

\bibitem{CCC}
Chien-Cheng Chang, Math.\  Comp.\ {\bf 49} 180 (1987)  523-542.

\bibitem{Petersen:1996by}
  W.~P.~Petersen,
  hep-lat/9602008.

\bibitem{Aarts:2012ft}
  G.~Aarts, F.~A.~James, J.~M.~Pawlowski, E.~Seiler, D.~Sexty and I.~-O.~Stamatescu,
  JHEP {\bf 1303} (2013) 073
  [arXiv:1212.5231 [hep-lat]].

\bibitem{arXiv:1001.3648}
  D.~Banerjee and S.~Chandrasekharan,
  Phys.\ Rev.\ D {\bf 81} (2010) 125007
  [arXiv:1001.3648 [hep-lat]].

\bibitem{arXiv:1104.2503}
  C.~Gattringer,
  Nucl.\ Phys.\ B {\bf 850} (2011) 242
  [arXiv:1104.2503 [hep-lat]].
  
\bibitem{Mercado:2012ue}
  Y.~D.~Mercado and C.~Gattringer,
  Nucl.\ Phys.\ B {\bf 862} (2012) 737
  [arXiv:1204.6074 [hep-lat]].

\bibitem{KW}
  F.~Karsch and H.~W.~Wyld,
  Phys.\ Rev.\ Lett.\  {\bf 55} (1985) 2242.
  
\bibitem{BGS}
  N.~Bilic, H.~Gausterer and S.~Sanielevici,
  Phys.\ Rev.\ D {\bf 37} (1988) 3684.

\bibitem{Greensite:2012xv}
  J.~Greensite and K.~Splittorff,
  Phys.\ Rev.\ D {\bf 86} (2012) 074501
  [arXiv:1206.1159 [hep-lat]].
  
\bibitem{Fromm:2011qi}
  M.~Fromm, J.~Langelage, S.~Lottini and O.~Philipsen,
  JHEP {\bf 1201} (2012) 042
  [arXiv:1111.4953 [hep-lat]].

\bibitem{Fromm:2012eb}
  M.~Fromm, J.~Langelage, S.~Lottini, M.~Neuman and O.~Philipsen,
  arXiv:1207.3005 [hep-lat].

\bibitem{Langfeld:2013kno}
  K.~Langfeld,
  arXiv:1302.1908 [hep-lat].

\bibitem{Campostrini:2000iw}
 M.~Campostrini, M.~Hasenbusch, A.~Pelissetto, P.~Rossi and E.~Vicari,
  Phys.\ Rev.\  B {\bf 63} (2001) 214503
  [cond-mat/0010360].
  
\bibitem{Batrouni:1985jn}
  G.~G.~Batrouni, G.~R.~Katz, A.~S.~Kronfeld, G.~P.~Lepage, B.~Svetitsky and K.~G.~Wilson,
  Phys.\ Rev.\ D {\bf 32} (1985) 2736.

\bibitem{bender}
  I.~Bender, T.~Hashimoto, F.~Karsch, V.~Linke, A.~Nakamura, M.~Plewnia, I.~O.~Stamatescu and W.~Wetzel,
  Nucl.\ Phys.\ Proc.\ Suppl.\  {\bf 26} (1992) 323.

\bibitem{feo}
  R.~De Pietri, A.~Feo, E.~Seiler and I.~-O.~Stamatescu,
  Phys.\ Rev.\ D {\bf 76} (2007) 114501
  [arXiv:0705.3420 [hep-lat]].



\end{thebibliography}
\end{document}